\documentclass[a4paper,11pt]{article}
\usepackage{jinstpub}
\usepackage{lineno}
\usepackage[separate-uncertainty=true]{siunitx}
\usepackage[nolist]{acronym}
\usepackage[noabbrev]{cleveref}
\usepackage{multirow}
\usepackage{caption}
\usepackage{subcaption}
\usepackage{tabularx,array,booktabs}
\newcolumntype{Y}{>{\raggedright\arraybackslash}X}   
\newcolumntype{P}[1]{>{\raggedright\arraybackslash}p{#1}}

\title{\boldmath Readout electronics for low occupancy High-Pressure Gas TPCs}

\author[a]{N. Khan}
\author[a]{, Y. Hua}
\author[a]{, I. Xiotidis}
\author[a]{, T. Alves}
\author[a]{, E. Atkin}
\author[b]{, G. Barker} 
\author[c]{, D. Barrow} 
\author[d]{, A. Booth} 
\author[a]{, J. Borg} 
\author[e]{, A. Bross} 
\author[f]{, M. F. Cicala} 
\author[d]{, L. Cremonesi} 
\author[g]{, A. Deisting} 
\author[c]{, K. Duffy} 
\author[h]{, R. Gran} 
\author[c]{, P. Green} 
\author[h]{, A. Habig} 
\author[l]{, M. Judah} 
\author[e]{, T. Junk} 
\author[i]{, A. Kaboth} 
\author[a]{, A. Klustová} 
\author[h]{, H. LeMoine} 
\author[j]{, A. D. Marino} 
\author[d,k]{, F. Martínez López} 
\author[k]{, T. Mohayai} 
\author[l]{, D. Naples} 
\author[f]{, R. Nichol} 
\author[a]{, D. Parker} 
\author[a]{, M. Pfaff} 
\author[e]{, J. Raaf} 
\author[e]{, P. Rubinov} 
\author[d]{, P. Singh} 
\author[g]{, A. Srivastava} 
\author[a,d]{, A. Waldron} 
\author[a,i]{, L. Warsame} 
\author[a,c]{, M. Wascko} 
\author[b,f]{, A. Wilkinson} 
\author[i]{, A. Ritchie-Yates} 
\author[a]{, and P. Dunne}
\affiliation[a]{Imperial College of London, London, UK}
\affiliation[b]{University of Warwick, Warwick, UK}
\affiliation[c]{University of Oxford, Oxford, UK} 
\affiliation[d]{Queen Mary University of London, London, UK} 
\affiliation[e]{Fermi National Accelerator Laboratory, Batavia, US}
\affiliation[f]{University College of London, London, UK} 
\affiliation[g]{Johannes Gutenberg-Universität Mainz, Mainz, GE}
\affiliation[h]{University of Minnesota Duluth, Duluth, US}
\affiliation[i]{Royal Holloway University of London, London, UK}
\affiliation[j]{University of Colorado Boulder, Boulder, US}
\affiliation[k]{Indiana University, Bloomington, US}
\affiliation[l]{University of Pittsburgh, Pittsburgh, US}

\emailAdd{i.xiotidis@imperial.ac.uk}

\abstract{
\acp{HPgTPC} have benefits such as low energy thresholds, magnetisability, and $4\pi$ acceptance, making them ideal for neutrino experiments such as \acs{DUNE}.
We present the design of an \acs{FPGA}-based solution optimised for \ac{ND-GAr}, which is part of the Phase-II more capable near detector for \acs{DUNE}. These electronics reduce the cost significantly compared to using collider readout electronics, which are typically designed for much higher occupancy and therefore, for example, need much larger numbers of FPGAs and power per channel. 
We demonstrate the performance of our electronics with the \ac{TOAD} at Fermilab in the US at a range of pressures and gas mixtures up to \SI{4.5}{barA}, reading out ${\sim}$\num{10000} channels from a \ac{MWPC}. The operation took place between April and July of 2024. 
We measure the noise characteristics of the system to be sufficiently low, and we identify sources of noise that can be further mitigated in the next iteration. We also note that the cooling scheme used in the test requires improvement before full-scale deployment. Despite these necessary improvements, we show that the system can fulfil the needs of a \ac{HPgTPC} for a fraction of the price of collider readout electronics. 
}

\keywords{Data acquisition concepts; Front-end electronics for detector readout; Neutrino detectors; Gaseous detectors}

\arxivnumber{2507.17425}

\begin{document}
\maketitle
\flushbottom

\begin{acronym}
\acro{TPC}{Time Projection Chamber}
\acro{FEC}{Front-End Card}
\acro{DFT}{Discrete Fourier Transformation}
\acro{OROC}{Outer Read-Out Chamber}
\acro{ADC}{Analogue to Digital Converter}
\acro{RMS}{Root Mean Square}
\acro{TOAD}{Teststand for an Overpressurised Argon Detector}
\acro{DUNE}{Deep Underground Neutrino Experiment}
\acro{HPgTPC}{High-Pressure Gas Time Projection Chamber}
\acro{ND-GAr}{Gaseous Argon Near Detector}
\acro{FPGA}{Field Programmable Gate Array}
\acro{DC/DC}{Direct Current to Direct Current}
\acro{IDC}{Insulation-Displacement Contact}
\acro{PAT}{Power, Aggregation and Timing}
\acro{ROC}{Read-Out Chamber}
\acro{DAQ}{Data Acquisition}
\acro{LDO}{Low-Dropout regulator}
\acro{PoE}{Power over Ethernet}
\acro{DC}{Direct Current}
\acro{HV}{High Voltage}
\acro{IROC}{Inner Read-Out Chamber}
\acro{SFP}{Small Form-Factor Pluggable}
\acro{CCM}{Control Configuration and Monitoring}
\acro{PCB}{Printed Circuit Board}
\acro{LVDS}{Low-Voltage Differential Signalling}
\acro{FSM}{Finite State Machine}
\acro{I2C}{Inter-Integrated Circuit}
\acro{SPI}{Serial Peripheral Interface}
\acro{PMbus}{Power-Management Bus}
\acro{UDP}{User Datagram Protocol}
\acro{TCP}{Transmission Control Protocol}
\acro{RR}{Round-Robin}
\acro{LHC}{Large Hadron Collider}
\acro{LAr}{Liquid Argon}
\acro{K-S}{Kolmogorov-Smirnov}
\acro{CLI}{Command Line Interface}
\acro{ASIC}{Application Specific Integrated Circuit}
\acro{MWPC}{Multi-Wire Proportional Chamber}
\acro{GEM}{Gas Electron Multiplier}
\acro{ECal}{Electromagnetic Calorimeter}
\acro{COTS}{Commercial Off the Shelf}
\acro{ND}{near detector}
\acro{ALICE}{A Large Ion Collider Experiment}
\end{acronym}

\section{Introduction}
\label{sec:intro}

Gas detectors offer greatly reduced energy thresholds, magnetisability, and 4$\pi$ acceptance, which are not currently achieved with other detector technologies \cite{abudDeepUndergroundNeutrino2021}. However, gases have much lower density than solid or liquid detectors, reducing the interaction rate for particles passing through. The advent of MW-power neutrino beams means that detectors with gas targets will, for the first time, be practical for long-baseline neutrino experiments. Despite these higher beam powers, a gas \ac{TPC} detector in these beams would still need to be pressurised to accumulate a sufficient number of interactions. 
 
Harnessing the capabilities of such a \acf{HPgTPC} for the \ac{DUNE} would provide important insights into neutrino-argon interactions and enable key measurements of the acceptance corrections of other detectors. As such, a gaseous argon \ac{TPC} surrounded by an \ac{ECal} has been recognised as crucial to the experiment's physics program by the recent US P5 process \cite{asaiExploringQuantumUniverse2023}, and the \ac{DUNE} collaboration's Phase II white paper \cite{NDGarRef}.

The detector's position resolution is a critical factor in achieving the required momentum resolution and detection thresholds for complex low-energy interactions with high multiplicity. The resolution obtainable is closely related to the number of readout channels that the detector can accommodate and the sampling frequency of those channels. Whilst the numbers are not final, instrumenting a \SI{5}{m} diameter \ac{TPC} sufficiently to achieve the desired low energy thresholds for the \ac{DUNE} would require $\mathcal{O}(\SI{100}{\kilo{}})-\mathcal{O}(\SI{1}{\mega{}})$ channels with sampling frequency $\mathcal{O}(\SI{10}{MHz})$\cite{abudDeepUndergroundNeutrino2021}. 
The development of readout electronics in the particle physics community has previously focused on the design and operation of high-throughput radiation-hard electronics required at the \ac{LHC}-based experiments \cite{CERN-LHCC-2013-020}. Readout electronics from sPHENIX \cite{sphenix} or the \ac{TPC} of \ac{ALICE} \cite{TheALICECollaboration_2008} require a high-specification \ac{FPGA} to readout groups of $\sim$10 front-end cards. Naively scaling such designs for neutrino experiments would be prohibitively expensive. This is due to the hundreds of thousands of channels required to meet \ac{DUNE}'s resolution needs. Hence, this approach would result in a total cost larger than possible, given budgetary constraints for the \ac{DUNE} near detector. In fact, the electronics were previously listed as the most expensive component for the \acf{ND-GAr} for \ac{DUNE}.

Since neutrino experiments have, by their nature, much lower occupancy and hence lower throughput than LHC experiments, a re-optimisation of the readout system enables significant cost savings whilst retaining physics performance. 

In the following sections we present the design of a scalable and highly re-configurable \ac{FPGA}-based solution optimised to be affordable for the \ac{ND-GAr} \ac{DUNE} Phase-II detector \cite{NDGarRef}. This system delivers significant cost-saving compared to the systems described above while showing no performance degradation in an \ac{HPgTPC} environment. A prototype has been produced and was then sent for testing to Fermilab as part of the \acf{TOAD}. Whilst \ac{DUNE} has not decided on an amplification stage for its \ac{HPgTPC}, a \ac{GEM}-based solution is the current frontrunner. However, \ac{TOAD} used a \acf{MWPC} from \ac{ALICE}, due to its availability at the time of the test.

\section{TOAD}
The \ac{TOAD} detector design and operation largely follow that of the UK high-pressure platform, described in \cite{Ritchie-Yates2023-cs}. The detector was sent to Fermilab, where it was commissioned, and the newly developed readout electronics were installed. These readout electronics interface with an \ac{ALICE} \ac{OROC}, meaning \ac{TOAD} was a full slice test of a \ac{HPgTPC}.

\subsection{ALICE Multi-Wire Proportional Chambers}
Given that the \ac{ALICE} \ac{TPC} \cite{Alessandro_2006, TheALICECollaboration_2008}  required an upgrade \cite{adolfssonUpgradeALICETPC2021}, the \acp{ROC} can be repurposed. The \ac{OROC} used in \ac{TOAD} was part of the \ac{ALICE} `test' \acp{ROC}, rather than a production \ac{ROC}. The specifications of the \ac{OROC} used are detailed at length in \cite{Ritchie-Yates2023-cs}.

The \acp{ROC} employ a typical wire plane schematic: an anode-wire grid above the pad plane, a cathode-wire grid and a gating grid to separate the drift volume from the readout. The pad plane and the three wire planes are all spaced \SI{3}{mm} apart. All wires run in the azimuthal direction (approximately vertically in the orientation we have the chamber) and have different separations in the various planes. The length of the shorter and longer parallel sides of the \ac{OROC} are \SI{45.8}{cm} and \SI{86.0}{cm}, respectively. The perpendicular length of the \ac{OROC} is \SI{114}{cm}. Voltages are applied to the anode and gating grid wires, while the cathode wires are held at ground. The pad plane consists of 9984 pads in total, of two sizes, which are summarised in \cref{tab:orocdimensions}. Here, ``inner'' refers to the rows closer to the shorter parallel side, whereas ``outer'' refers to rows closer to the longer side of the \ac{OROC}. The difference in pad dimensions is due to the trapezoidal shape of the \ac{ROC}. The readout electronics must interface with this \ac{ROC}. Therefore, groups of 21 or 22 pads are read out via a socket on the back of the \ac{OROC}. With the sockets grouped in threes, our \acp{FEC} connect directly to these to read out a total of 64 channels each.
\begin{table}[h]
    \centering
    \caption{Dimensionality of the pad plane on the OROC.}
    \label{tab:orocdimensions}
    \begin{tabular}{c|c|c|c}
        & \textbf{Pad Dimensions (mm$^{2}$)} & \textbf{Pad Rows} & \textbf{Number of Pads} \\
        \hline
        \hline
        Outer Pads & 6 $\times$ 15 & 32 & 4032 \\
        Inner Pads & 6 $\times$ 10 & 64 & 5952 \\
        Total & - & 96 & 9984 \\
        \hline
    \end{tabular}
\end{table}

\subsection{Pressure vessel}
\begin{figure}[h]
    \centering
    \includegraphics[width=0.6\textwidth]{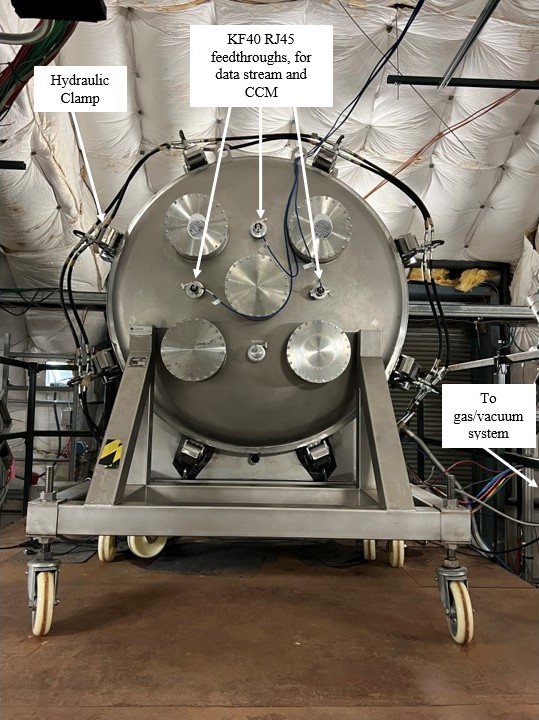}
    \caption{The \ac{TOAD} pressure vessel in the test environment at Fermilab. Around the vessel, the hydraulic pressure clamps can be seen. In the centre, there is access to five flanges, which are unused in \ac{TOAD} but previously used for mounting an optical readout system. The KF40 feedthroughs that are used for data readout and \ac{CCM} via the blue cables are labelled. Various other KF40 and KF25 flanges on the vessel provide access for the different subsystems, but cannot be seen in this image. The gas and evacuation system is to the right of the vessel.}
    \label{fig:toad_vesselfront}
\end{figure}

\ac{TOAD} utilises the same pressure vessel described in \cite{Ritchie-Yates2023-cs, instruments5020022}, which allows the testing of the readout electronics with an \ac{ALICE} \ac{OROC} at pressures up to \SI{4}{barG} ($\sim$\SI{5}{barA}). The vessel has an inner diameter of \SI{140}{cm} and an outer diameter of \SI{142}{cm}, made from \SI{304}{L} Stainless Steel. The vessel has domed ends, giving a total volume of \SI{1.472}{m^3}. The vessel can be opened by removing one of the domed faces. An image of the vessel in the test environment is shown in \cref{fig:toad_vesselfront}. Further details of the vessel properties can be found in \cite{Ritchie-Yates2023-cs, instruments5020022}. The way in which the feedthroughs were utilised for \ac{TOAD} differs from the original setup detailed in \cite{Ritchie-Yates2023-cs}. In addition, the vacuum system was replaced. There are various KF25 and KF40 flanges on the vessel, which are used for the \ac{HV}, gas, vacuum and readout-electronics systems feedthroughs. The gas and evacuation system is connected to the vessel so that the vessel can be evacuated to $\mathcal{O}\left( \SI{e-5}{barA}\right)$ to ensure the purity of gas in the detector. The pump system was upgraded to an Edwards Turbo Cart, consisting of a backing pump, a turbo pump and a controller. The gas system allowed the vessel to be filled with up to \SI{5}{barA} of gas mixtures, with the gas mix controlled by filling to a given partial pressure of gas from up to four different gas cylinders connected to the system. For the run described in this paper, three input gases were mixed: Ar-CH$_{4}$ with a molar ratio of 92:8, pure Ar, and pure N$_{2}$.

The \ac{TPC} is the same as that described in \cite{Ritchie-Yates2023-cs}. It utilises a \ac{HV} drift cathode, a field cage, a terminating plate, and the \ac{OROC}. There are three \ac{HV} feedthroughs on the vessel, which are used to apply voltages to the drift cathode, anode and gating grid. Other feedthroughs are also used to ensure the equipment inside the vessel is all safely grounded. The cathode and \ac{OROC} anode were previously tested with voltages up to \SI{16}{kV} and \SI{3}{kV} respectively, with the maximum achievable voltage varying with the gas mixture used \cite{Ritchie-Yates2023-cs}. The results shown in this paper use various configurations of \ac{HV}, gas mixtures and pressures. These are summarised in \cref{tab:gasmixtures}. The measurements outlined in this paper are noise characterisation studies, and so the particle tracking ability of the \ac{TPC} is not employed.

\begin{table}[h]
    \centering
    \caption{Different operational configurations used for data shown in the Results section.}
    \label{tab:gasmixtures}
    \begin{tabular}{c|c|c|c|c|c}
        \textbf{Mode} &  \textbf{Gas Mixture} & \textbf{Pressure (barA)} &\textbf{Anode (V)} & \textbf{Gating Grid (V)} & \textbf{Cathode (V)}\\
        \hline
        \hline
         1 & Air & 1.0 & 0 & 0 & 0 \\
         2 & Ar:CH$_{4}$, 96:4 & 4.5 & 0 & 0 & 0 \\
         3 & Ar:CH$_{4}$, 98:2 & 3.0 & 1500 & \num{-100} & 0 \\
         4 & Ar:CH$_{4}$, 98:2 & 3.0 & 1500 & \num{-100} & \num{-16000} \\
         \hline
    \end{tabular}
\end{table}

In addition to the connections described above, the readout electronics were powered and streamed data through KF40 feedthroughs on the vessel door. One feedthrough was used to provide power, and three RJ45 feedthroughs are used to control, configure, monitor and stream data to the readout electronics. These were connected directly to a \ac{COTS} network switch, which was, in turn, connected to two DELL PowerEdge servers: one running \ac{CCM} software, and the other \ac{DAQ} software -- both of which are outlined in detail in \cref{SoftwareDevelopment}.

\section{Readout setup}

A key challenge for future neutrino experiments is the harsh conditions in which the electronics must operate. In contrast to collider experiments, those conditions are not due to extensive radiation but instead factors such as cryogenic temperatures present in \ac{LAr}-based detectors and high pressure, such as in an \ac{HPgTPC} based detector. 
In the case of the high-pressure detectors, an important consideration is the number of feedthroughs required to power and communicate with the electronics, so there is a need to aggregate data inside the vessel. Whilst power dissipation and thermal management present additional challenges, they can be addressed using water cooling through a small number of vessel penetrations. Hence, a readout system will need to optimise the number of cables required to stream data from the large number of analogue readout channels. In addition, based on the physics requirements for the \ac{DUNE} \ac{ND} complex outlined in the Conceptual Design Report \cite{abudDeepUndergroundNeutrino2021}, the proposed readout system should not constrain or limit the advantages provided by the installation of a magnetised high-pressure gaseous \ac{TPC} in the complex. Such requirements foresee enhanced tracking capabilities in comparison to the \ac{LAr}-based detector counterpart for more accurate event classification. A summary of the requirements taken into account can be seen in \cref{tab:NDGArRequirements} with the equivalent design choice and the technology implementation. Obviously, choices like the number of links aggregated into a single \ac{FPGA} are driven based on the capabilities of the package as well as the cost, making it an optimisation process based on physics versus cost.

\begin{table}[h]
  \centering
  \setlength{\tabcolsep}{6pt}        
  \renewcommand{\arraystretch}{1.15}  

  \begin{tabularx}{\linewidth}{P{3.6cm}|Y|Y}
    \textbf{Requirement} & \textbf{Design choice} & \textbf{Implementation}\\
    \hline\hline
    Detection threshold ($\mathcal{O}$(MeV) protons) & High-Pressure \ac{TPC} & Reduced feedthroughs \\
    \multirow{2}{8em}{Track separation ($1-\SI{5}{cm}$)} & \ac{ADC} sampling       & \SI{20}{MHz} \\
                                         & Num.\ of detector channels & $\mathcal{O}$(700k) \\
    Low event rates (\SI{1}{Hz})              & Gigabit readout         & Kintex Ultrascale \ac{FPGA} \\
    \hline
  \end{tabularx}

  \caption{Physics requirements and expected timing and throughput values for \ac{ND-GAr} as described in \cite{abudDeepUndergroundNeutrino2021} with the corresponding choices in the design and solutions applied to those decisions. The requirement on the track length resolution arises from the ALICE \ac{TPC} capabilities, which should suffice for \ac{ND-GAr} but will be refined in the future.}
  \label{tab:NDGArRequirements}
\end{table}

\begin{figure}
    \centering
    \includegraphics[width=0.8\linewidth]{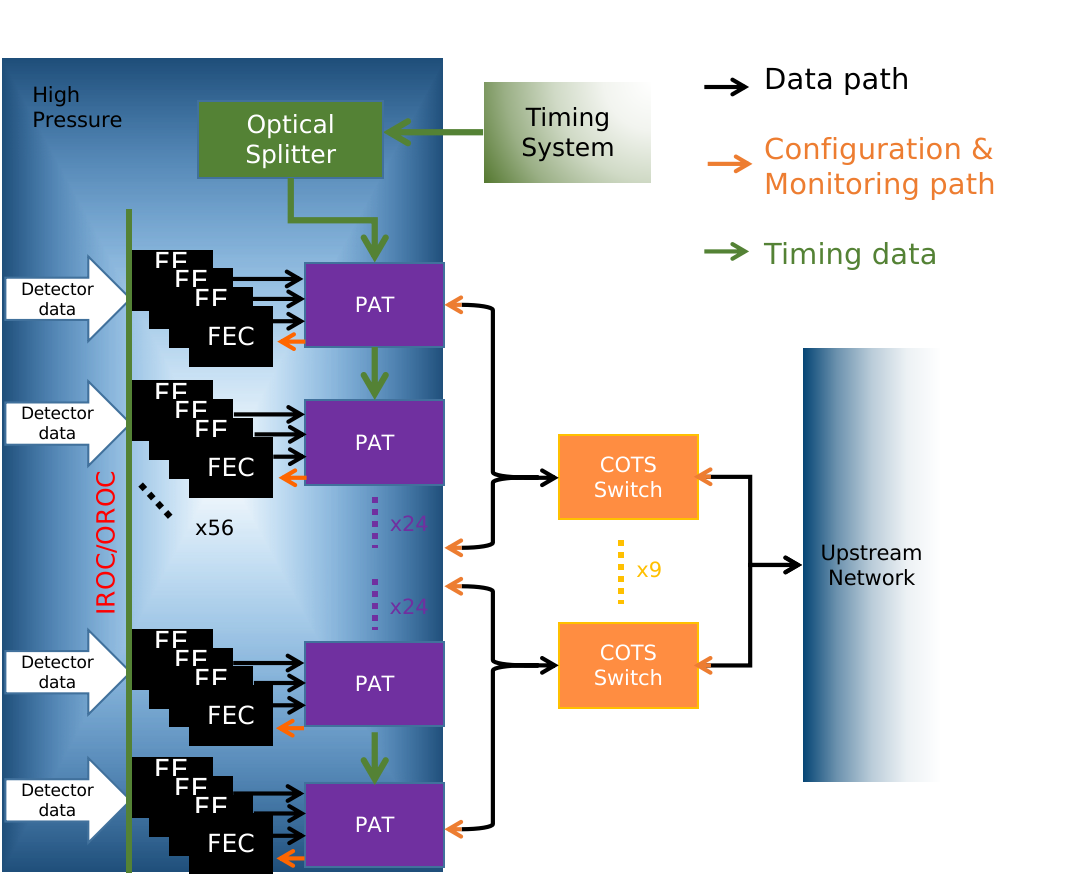}
    \caption{\ac{ND-GAr} readout scheme with direct power provided to all \ac{DAQ} cards and explicit timing signal distribution. The vertical, dark green line represents the \ac{ALICE} Inner/Outer Readout Chamber. The orange boxes represent \ac{COTS} network switches.}
    \label{fig:NDGarReadout}
\end{figure}

The proposed solution outlined in this paper uses a fast-sampling \ac{FEC} with zero-suppression capability and a custom \ac{FPGA}-based \ac{PAT} card, both residing in-situ within the high-pressure volume. Data are then streamed over a standard Ethernet-based network to the \ac{DUNE}~\ac{DAQ} software operating on standard servers, avoiding extensive requirements within the pressurised environment. The block diagram of the proposed readout system is shown in \cref{fig:NDGarReadout}.

Compared to a collider experiment, the channels in neutrino experiments have very low occupancy (in the similarly sized \ac{ALICE}, collisions occur at $\sim$\SI{35}{kHz} and each generate thousands of tracks, whereas in \ac{DUNE}, beam spills happen at $\sim$\SI{1}{Hz} and generate only tens of tracks per spill). This difference means far less processing is required as there are far fewer above-threshold hits. Therefore, the system can have fewer FPGAs per channel (1 per \num{3584} channels with the particular \ac{FEC}s in this design) and use much lower specification FPGAs as we stream out all hits to a software-based DAQ. Comparisons with the \ac{ALICE} and sPHENIX readout electronics are shown in \cref{tab:SystemComparison}.

\begin{table}[h]
    \centering
    \begin{tabular}{c|c|c|c}
         \textbf{System} & \textbf{Event Rate} & \textbf{Sampling Rate} & \textbf{Channels/board} \\
         \hline
         \hline
         sPHENIX & \SI{15}{kHz} & \SI{60}{MHz} & 192 \\
         ALICE & \SI{35}{kHz} & \SI{10}{MHz} & Vary per detector region \\
         ND-GAr & \SI{1}{Hz} & \SI{20}{MHz} & 3584\\
         \hline
    \end{tabular}
    \caption{Comparison in terms of event rates, sampling rates and channels per readout card among the sPHENIX readout system, the ALICE calorimeter readout and the ND-GAr readout.}
    \label{tab:SystemComparison}
\end{table}

Furthermore, the lower data rates from the FEC allow the native transceivers on the \ac{ADC} \acp{ASIC} to be used to communicate with the \ac{PAT}, so we also do not need the additional transceiver modules that are required for higher occupancy detectors \cite{adolfssonUpgradeALICETPC2021}. These differences result in a much lower cost to instrument the hundreds of thousands of channels necessary for a neutrino \ac{HPgTPC}. 

\subsection{Hardware layout}

As described, our system requires two types of custom hardware boards. The first type are the \ac{FEC} boards, which host the digitiser \acp{ASIC} and connect to the detector itself. The second type are the \ac{PAT} boards, which utilise an \ac{FPGA} device for interfacing and aggregating the data for a multitude of \ac{FEC}s as well as providing them with power and distributing clock signals. Both the \acp{FEC} and the \acp{PAT} are custom-designed, with the \ac{FEC}s designed for the \ac{TOAD} test beam detector and its \ac{ALICE} \ac{OROC}. The flexibility of the \ac{FPGA}-based design of the \ac{PAT} cards means that the \ac{PAT} can be reused even if the \ac{FEC}s change, so, whilst designed for \ac{TOAD}, this board could be used with minimal modifications for the final \ac{ND-GAr} detector or other detectors with similar data volume and occupancy. In designing the hardware, we avoided components with known air gaps, \textit{e.g.} lidded \ac{FPGA} packages. However, as there is no standard for electronics operation in the \SI{10}{bar} regime, a goal of \ac{TOAD} is to test the operation of components at high pressures.

\subsubsection{Front-End Card}
\ac{GEM}s are currently the leading candidate for the charge readout for \ac{ND-GAr}. The SAMPA \ac{ASIC} used for this readout system was developed by the \ac{ALICE} collaboration for the \ac{GEM} upgrade of their \ac{TPC} \cite{adolfssonUpgradeALICETPC2021}. This upgraded detector has similar characteristics to those needed for \ac{ND-GAr}, so SAMPA is expected to meet the requirements of \ac{ND-GAr}. The parameters of the SAMPA measured by \ac{ALICE} can be seen in \cref{tab:SAMPAParameters} \cite{adolfssonSAMPAChipNew2017, adolfssonUpgradeALICETPC2021}. The electronics system described here is, therefore, designed around deploying the SAMPA \ac{ASIC} \cite{adolfssonSAMPAChipNew2017} used by \ac{ALICE}, in a high-pressure environment, and at a much lower cost than in the full \ac{ALICE} system.   

\begin{figure}
    \centering
    \includegraphics[width=0.5\linewidth]{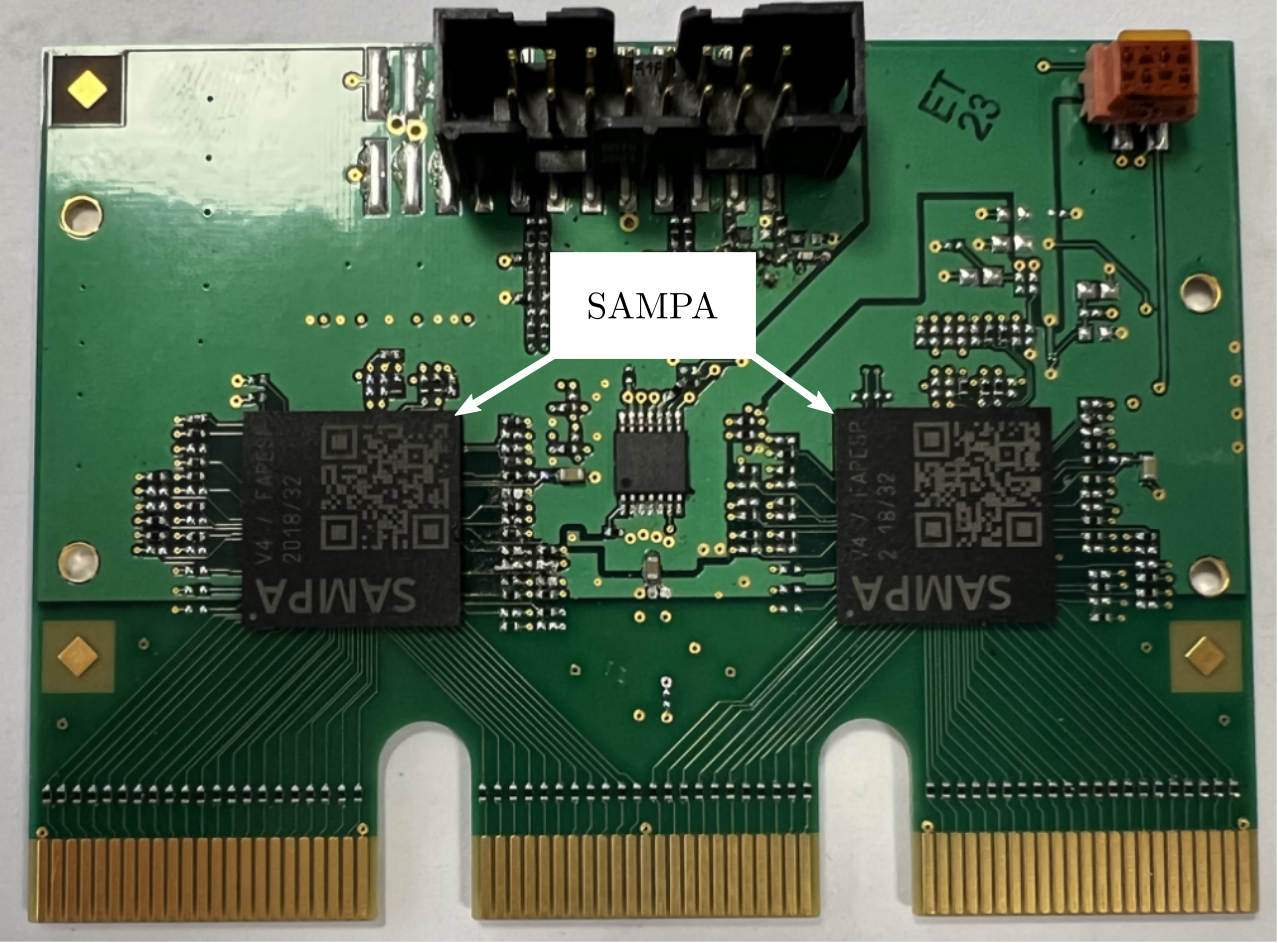}
    \caption{\ac{FEC} board hosting two daisy-chained SAMPA chips designed for the \ac{TOAD} \ac{OROC} based test beam readout.}
    \label{fig:SAMPAPic}
\end{figure}

\begin{table}[h]
    \centering
    \caption{SAMPA reference parameters measured for the \ac{ALICE} \ac{TPC} upgrade with \acp{GEM} \cite{adolfssonSAMPAChipNew2017, adolfssonUpgradeALICETPC2021}.}
    \label{tab:SAMPAParameters}
    \begin{tabular}{c|c}
        \textbf{Specification} & \textbf{GEM} \\
        \hline \hline
        Supply Voltage & \SI{1.25}{V} \\
        Detector Capacitance & 40-\SI{80}{pF} \\
        Peaking Time & \SI{300}{ns} \\
        Shaping Order & 4th \\
        Equivalent Noise & <\num{1600}e Cd=\SI{80}{pF} \\
        Sampling Rate & \SI{10}{MHz}, up to \SI{20}{MHz} \\
        Total Power (Idle)& $<\SI{15}{mW/Ch}$ \\
        \hline
    \end{tabular}
\end{table}

In addition, using an \ac{ASIC} that has been proven to work at scale for another particle physics detector minimises risks and allows us to benefit from the extensive studies of their performance and limitations that have already been performed.

Another important reason for the choice of the SAMPA chips is their ability to sample at \SI{20}{MHz} with fast \acp{ADC}. Gas detectors have about an order of magnitude faster electron drift velocities than liquid; hence, this high sample frequency is necessary. 

For \ac{TOAD}, the \ac{ALICE} \ac{MWPC} has a group of three connectors per slot, which correspond to a total of 64 channels. Given that the SAMPA \ac{ASIC} has only 32 channels, the \ac{FEC} designed for \ac{TOAD}, which can be seen in \cref{fig:SAMPAPic}, has three connectors and hosts two SAMPAs to capture all 64 channels. The final \ac{ND-GAr} amplification stages may have a different number of channels per connector; however, the more SAMPAs per \ac{FEC}, the lower the cost of the system, so designing for two SAMPAs represents a reasonable worst-case cost. To minimise the required readout channels connecting to the \ac{PAT} board, the two SAMPAs on each \ac{FEC} are daisy-chained, as the low occupancy of neutrino experiments does not require the full bandwidth of the SAMPA readout link. However, as this is a prototype detector, the layout of the \ac{FEC} has been designed to allow each SAMPA to be read out individually with minimal changes to the fitted components and no re-design of the \ac{PCB}.

For \ac{TOAD}, the \ac{FEC}s were operated in the configuration shown in \cref{tab:TOADSampas}. This configuration gives an expected current draw of \SI{800}{mA}/channel with the \ac{FEC} operating at full capacity. It is, however, not expected that the \ac{FEC} will have to operate at full power due to low occupancy and zero suppression. 
Whilst the design of the \ac{FEC} is specific to \ac{TOAD}, its operation at \ac{TOAD} showcases the expected performance of SAMPA more generally in the type of environment that is required for \acp{HPgTPC}, which has not been previously tested.

\begin{table}[h]
    \centering
    \caption{Operation parameters for SAMPA \ac{ASIC}.}
    \label{tab:TOADSampas}
    \begin{tabular}{c|c}
        \textbf{Parameter} & \textbf{Value} \\
        \hline \hline
        Supply Voltage & \SI{1.25}{V} \\
        Reference Clock & \SI{160}{MHz} \\
        Sampling Frequency & \SI{20}{MHz}\\
        Operation Mode & Streaming/Trigger \\
        \hline
    \end{tabular}
\end{table}

\subsubsection{Power Aggregation and Timing cards}

\begin{figure}[h]
    \centering
    \includegraphics[width=0.5\linewidth]{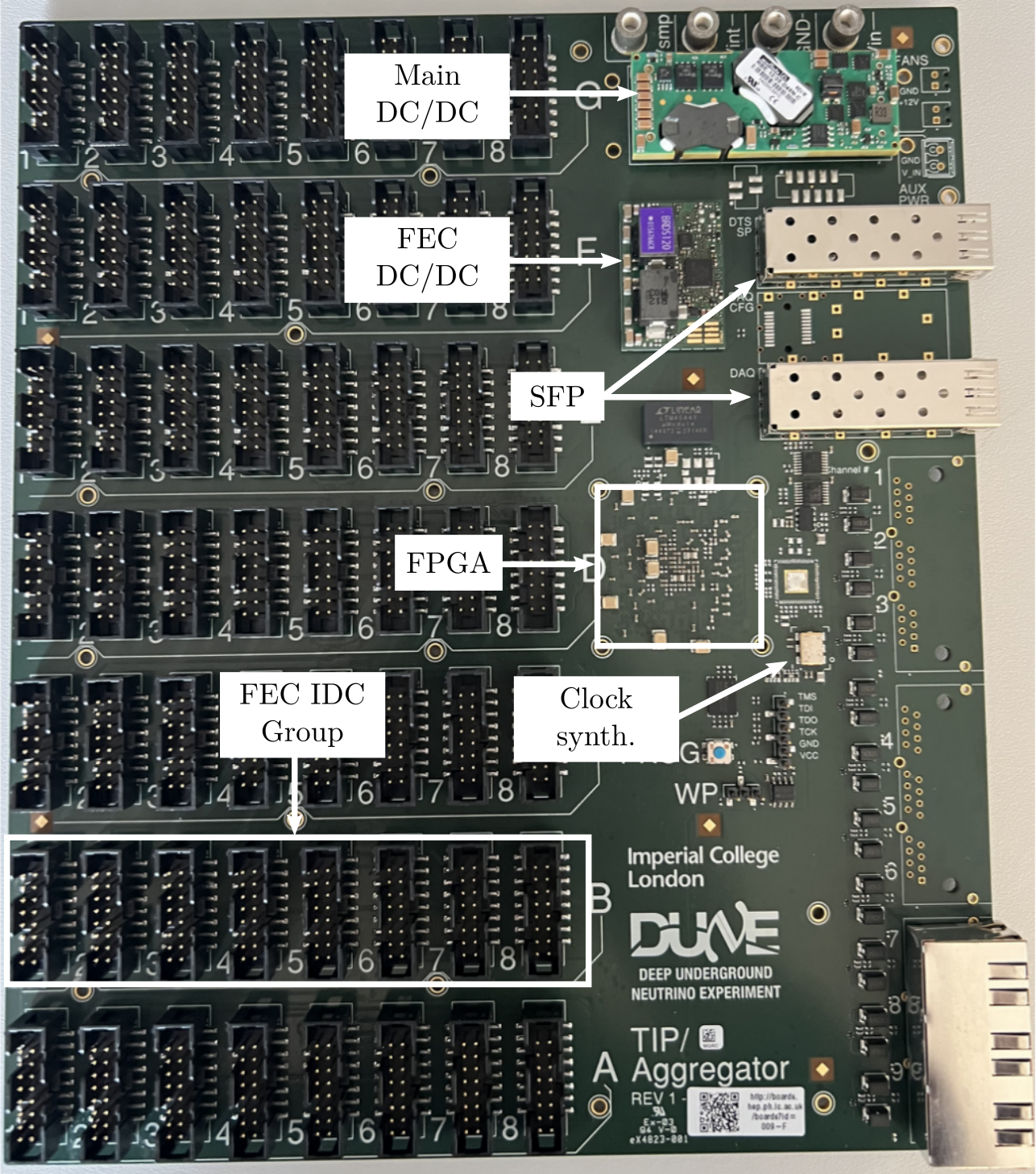}
    \caption{\ac{PAT} card top-level image, on the top the two \ac{DC/DC} converters can be seen, one dedicated for the main voltage translation (Main \ac{DC/DC}) and the second one for the \ac{FEC} power (\ac{FEC} \ac{DC/DC}). The \ac{FPGA} position is on the back side of the board at the position indicated, the \ac{FEC} communication is organised in groups of \ac{IDC} connectors shown with the white box on the left. Finally, there are four placeholders for clock synthesizers, however, only two are mounted, with one of them indicated by the white arrow on the right.}
    \label{fig:PATCard}
\end{figure}

The \ac{PAT} card's primary function is to interface between the \ac{FEC} cards and a standard Ethernet network, where data can be streamed to servers for data selection and storage. The prototype, designed at Imperial College London, utilises the advantages of \ac{FPGA}s as the main data processing unit for the \ac{PAT}. An image of the prototype board can be seen in \cref{fig:PATCard}. 

\begin{figure}
    \centering
    \includegraphics[width=1.\linewidth]{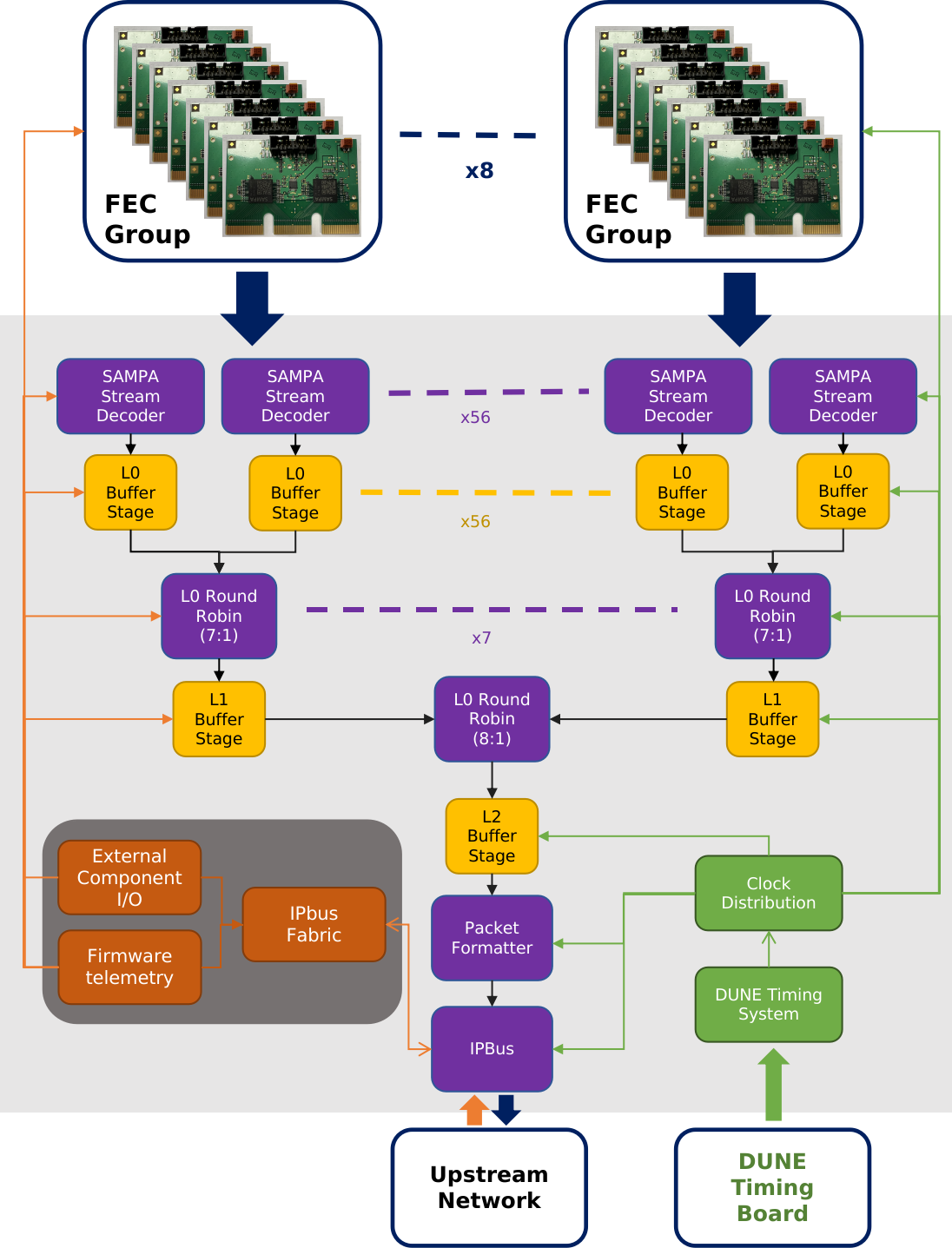}
    \caption{Diagram of the firmware schematic. The black lines show the data path, through successive levels of buffering (L0, L1, L2), format conversions and the \ac{RR} data flow management scheme. The green lines show the clock distribution (derived from the \ac{DUNE} Timing System). The orange lines show telemetry signals to/from internal/external components and the \acs{CCM} system. The thicker arrows denote serial communication to other boards or switches in the system.}
    \label{fig:FWBlockDiagram}
\end{figure}

Data aggregation is a primary requirement that drives the design of the \ac{PAT}. The cards are designed to take inputs from \num{56} \acp{FEC} each streamed over a differential pair into the \ac{FPGA} package. The received data payloads from the many \acp{FEC} are then extracted and propagated through a total of \num{64} buffers using \ac{RR} logic to aggregate them into a single output stream. This output data stream is then wrapped into an Ethernet-based protocol and sent over the \ac{FPGA} serial transceivers to the \ac{COTS} network switch. The full layout can be seen in \cref{fig:FWBlockDiagram}. Due to the low occupancy of the detector, the aggregation factor of 56 to 1 can be achieved easily within the \ac{FPGA} fabric. 

As well as data aggregation, the powering and timing roles of the \ac{PAT} are also crucial. For powering, we investigated two possible solutions, for which a summary of key metrics is provided in \cref{tab:FeedthroughCount}. The first solution provides \SI{48}{V} to each \ac{PAT} via dedicated terminals on the board. The \ac{FPGA} \ac{DC/DC} converter then outputs \SI{12}{V} which is further fed into the \ac{FEC} \ac{DC/DC} converter to be transformed to the \SI{1.8}{V} needed by the \ac{FEC} on-board \ac{LDO}. The advantage of this scheme is that each \ac{PAT} card receives a dedicated power line, minimising potential interference with signal traces on the board. The disadvantage is that \ac{ND-GAr} will need dedicated power feedthroughs in the pressure vessel, which increases the overall feedthrough count. 

\begin{figure}
    \centering 
    \includegraphics[width=0.8\linewidth]{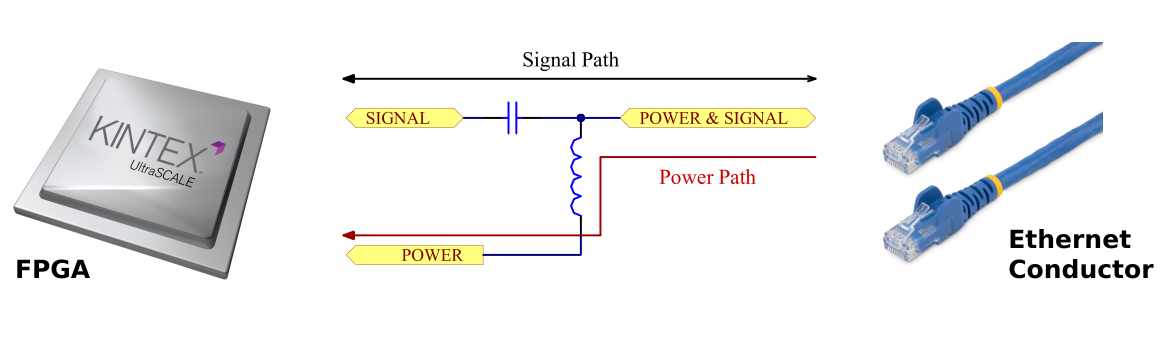}
    \caption{Circuit diagram for an alternate powering scheme where data and power are provided over the same cable. Power is injected by biasing the POWER \& SIGNAL line to \SI{48}{V} and thereby also the subsequent Ethernet Conductor (seen with red arrow: Power Path). At the \ac{FPGA}, the capacitor causes the SIGNAL pins to only see the alternating component of the voltage and thereby removes the \ac{DC} bias. Finally, the inductor acts as an open circuit switch, ensuring that signal doesn't leak towards the POWER node.}
    \label{fig:PWRInjection}
\end{figure}

The second `power over serial' powering scheme is inspired by \ac{PoE}, and aims to allow \ac{PAT} cards placed within the pressure vessel to receive power over the incoming network cable. The circuit diagram used can be seen in \cref{fig:PWRInjection}. However, some modifications to the normal \ac{PoE} standard are required due to the intended environment. \ac{PoE} works by injecting a \ac{DC} component on top of the alternating data component using transformers. Copying the same methodology for a magnetised detector wouldn't be viable, as transformer performance may be impacted by the surrounding magnetic field. Therefore, an alternative system was developed using air-cored inductors to extract the differential signal from the bias voltage. 
Such a scheme provides the following two advantages. Primarily, the number of feedthroughs and cables required for the complete detector readout would be reduced compared to the direct power solution. A second major difference with this option is that a board that is outside of the pressure vessel is required to receive these power plus data signals. This second board would allow additional buffering and processing to be done on the \ac{FPGA} if necessary. In the explored design, both the \ac{PAT} boards inside and outside of the pressure vessel were based on the same board design with minor differences in the fitted components. 

The implementation on an early prototype for the \ac{PAT} cards was found not to give sufficient signal stability, so further optimisations are required to achieve a production-level system with this powering option. This work is underway with a new dedicated hardware board allowing modifications to both the power injection and the power decoupling circuits for further analysis. However, as it has not yet converged, we therefore chose the first solution (direct power) for the board design discussed here. 

In addition, a study, summarised in \cref{tab:FeedthroughCount}, was done to estimate the number of feedthroughs needed for both options and also a third potential solution which would use optical fibres to stream out the data and a separate power line for each \ac{PAT}. This option allows several \ac{PAT} cards to share one output feedthrough as optical fibres are smaller in form factor than the electrical connectors needed.

\begin{table}[h]
    \centering
    \caption{Feedthrough and component count for different readout schemes when using the \ac{PAT}-based system for an $\sim$\num{700000} channel detector.}
    \label{tab:FeedthroughCount}

    \begin{tabular}{c|c|c|c}
        \textbf{Component} & \textbf{Power over Serial} & \multicolumn{2}{c}{\textbf{Direct Power}}\\
        \hline
        Readout Type& Electrical & Optical & Electrical \\
        \hline
        \hline
        \# \ac{PAT} (In-Pressure) & 196 & 196 & 196\\
        \# \ac{PAT} (Out-Pressure) & 22 & 0 & 0 \\
        \# \ac{FPGA} & 218 & 196 & 196 \\
        \# Feedthrough & 196 & 76 & 206 \\
        \# Net. Switch Eth. Ports & 22 & 196 & 196 \\
        \hline
    \end{tabular}

\end{table}
Decisions on the power distribution of the system also have an impact on the timing. For the Power over Serial method, timing can be provided to the upstream \ac{PAT} card that is outside the pressure vessel and all the cards within the vessel can then be synchronised by the serial transceiver recovered clock. In the case of Direct Power, the data stream is separate from power and is running through commercial \ac{SFP} modules. We therefore require a separate dedicated input for clock distribution. The \ac{PAT} cards are designed to be able to interface with the \ac{DUNE} timing system \cite{DTSRef} over an optical fibre. In order to implement this, extra optical feedthroughs are required to receive all the timing fibres into the pressure vessel. The number of feedthroughs can be mitigated by fanning out the signal inside the vessel, but there are limits to the number of boards that can be supplied from any one fibre. In both cases, the clock received at the \ac{PAT} cards in the pressure vessel is transmitted from the \ac{FPGA} to a dedicated clock synthesiser for cleaning and then further transmitted to the \ac{FEC}s. 

Furthermore, distribution of the \ac{FEC} clock, synchronisation, and slow control signals happen, depending on the use case, via fan-out chips or multiplexers. Given that the designed readout is synchronous to the \ac{DUNE} clock, all the \ac{FEC} cards are expected to receive the same copy of the clock synchronisation signal. For this reason, an organisation of eight \acp{FEC} in seven groups has been designed. The advantage of this method relies on the layout design of the \ac{PCB}. The chosen \ac{FPGA} can drive seven individual groups with the same clock synchronisation signal pulse. Each of these seven individual groups is then copied in phase to all eight \acp{FEC} connected to that pin. The main reason for deciding to fan-out 1:8 was to ensure that the clock quality provided to all the \acp{FEC} wasn't impacted by the fan-out \ac{ASIC}. Equivalently, for the slow control interface, a series of multiplexers is implemented, allowing individual access to all the \ac{FEC}s, providing enhanced telemetry of the \ac{FEC} operation.

\begin{figure}[h]
    \centering
    \includegraphics[width=0.6\textwidth]{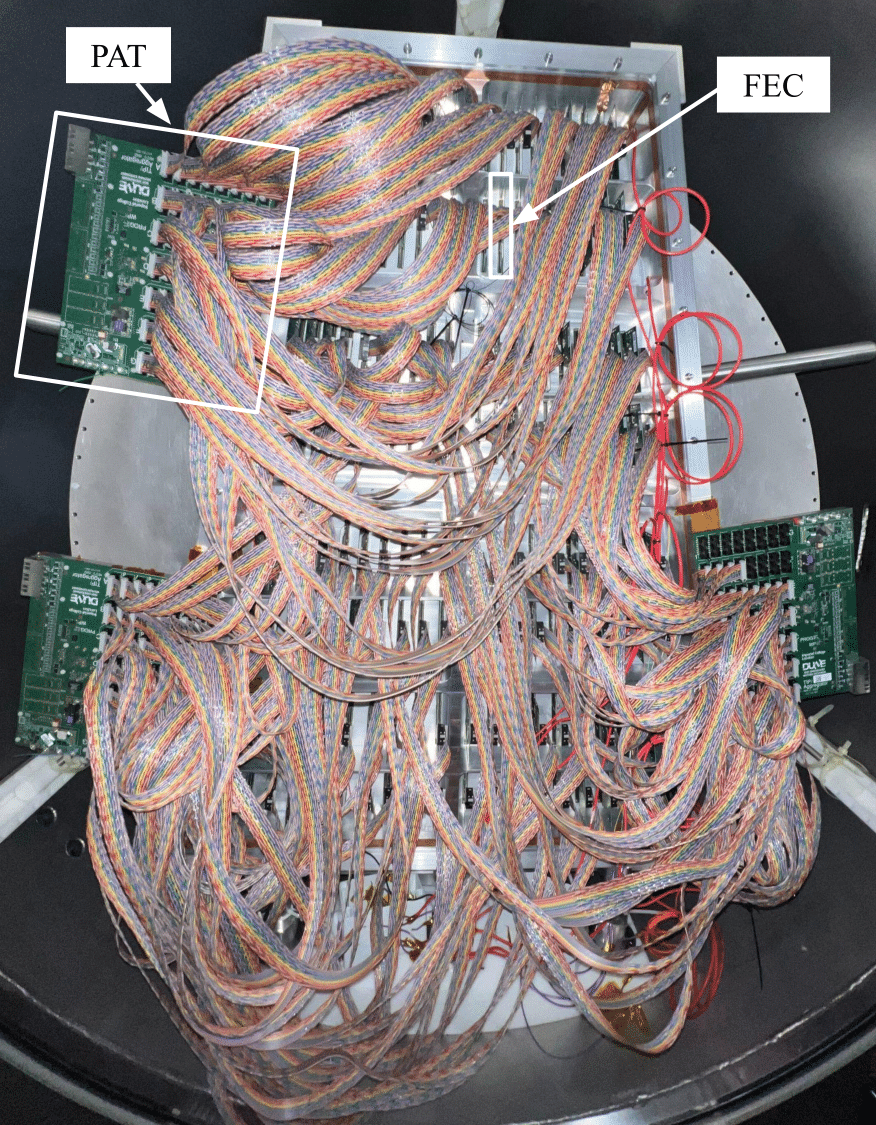}
    \caption{The readout side of the \ac{OROC}, which has 156 \ac{FEC}s connected to 3 \ac{PAT} boards via ribbon cables. These cables are either \SI{0.5}{m} or \SI{1.0}{m} in length. The \ac{PAT} boards are mounted on aluminium plates and secured to the \ac{OROC} frame. Each \ac{FEC} has 3 connectors to the \ac{OROC}, where each connector reads out groups of 21 or 22 pads. In this way, each \ac{FEC} reads out the charge signal on 64 channels/pads, and so all 9984 channels of the \ac{OROC} can be read out. A \ac{PAT} and \ac{FEC} have been highlighted in this image.}
    \label{fig:electronicsinvessel}
\end{figure}

Finally, the \ac{PAT} cards are mounted at three positions on the \ac{OROC}, in such a way that when 18 \acp{OROC} are mounted into a circle, the \ac{PAT} cards are shared between neighbouring \acp{OROC}. In this way, the cable lengths required between \ac{FEC}s and \ac{PAT} cards are minimised, and so, these connections are optimised. The \ac{PAT} cards are mounted on aluminium plates, which, through thermal contact with the \ac{ROC}, provide some amount of passive cooling. An image of the \acp{FEC} and \acp{PAT} connected to the \ac{OROC} inside the vessel is shown in \cref{fig:electronicsinvessel}.

\subsection{Firmware design}
The \ac{PAT} card's main processing unit is a Kintex Ultrascale \ac{FPGA} package. This device was selected as it provides the required number of serial transceivers to accommodate both readout schemes for \ac{ND-GAr} whilst satisfying the remaining requirements. The power consumption (based on the AMD power estimate tool) is within the nominal limits for operation in gases at atmospheric pressure or above when actively cooled. Finally, the availability of this package at low cost through CERN and the extensive experience of using it that has been built up at CERN make this \ac{FPGA} a favourable choice given the cost-driven nature of the design. 

In order to take advantage of the \ac{FPGA}, dedicated firmware had to be developed that allowed data to be reliably and efficiently streamed. Here, we discuss the firmware solution for direct powering, as this was the firmware that was eventually tested with the \ac{TOAD} test beam. 
The design has two primary functions. First, the data flow, \textit{i.e.} how \ac{FEC} data are reliably received into the software. The second is \acf{CCM} operations. A schematic representation of the firmware design can be seen in \cref{fig:FWBlockDiagram}.

\subsubsection{Data flow}\label{sec:dataflow}

The main objective of the data flow firmware is to reliably stream the \ac{FEC} data into a server running the \ac{DUNE} \ac{DAQ} software \cite{a.abedabudTriggerDataAcQuisition2023}. The input to this firmware is the data stream from the daisy-chained SAMPAs, which is fed into the \ac{FPGA} pins via a \ac{LVDS} stream. The SAMPAs stream first a $50b$ header which includes important information, like the number of $10b$ samples that will be streamed, the time of arrival information, the fired channel number, etc. After the header, the payload follows which is made up of N $10b$ words, where N is defined in the header.

The data stream is then decoded using dedicated decoder firmware based on a \ac{FSM} that can handle all the different operational modes of the SAMPA. The extracted payload, along with some of the header information (channel number, timestamp, and number of clusters), is converted into a $32b$ stream that is subsequently sent through a \ac{RR} based buffering system. The output Ethernet sample additionally contains a single $32b$ header with some of the header information received from the SAMPA header, as well as the $64b$ \ac{DUNE} timestamp for data alignment.

The firmware implements three options for the streaming protocol used to do this. The first option, which is the one we used for all data shown here, uses the standard IPbus library over \ac{UDP} \cite{IPbusCite}. This option provides a stream of a given number of words on request from the \ac{DAQ} software.

The second option interfaces the SiTCP suite from BeeBeans \cite{SiTCPRef} with IPbus. SiTCP is a lightweight interface that provides streaming capabilities over \ac{TCP}. We developed an interface between the standard pull-request based IPbus firmware and the SiTCP IP core, thereby providing a \ac{TCP}-based interface for the IPbus suite. While SiTCP includes its own \ac{UDP}-based interface for \ac{CCM}, it does not have enough address space for our application.

The third option was used for the power over serial powering mode investigations. The number of serial links on the \ac{PAT} only allows for a single hardware interface between the in-pressure and out-of-pressure boards for both data and \ac{CCM} and hence the SiTCP solution can be used in the out-of-pressure electronics. The link between the in-pressure and out-of-pressure electronics still cannot use the SiTCP solution, so it is implemented as an Aurora-based stream interface. In this operation mode, IPbus is not used at all, and the received data are converted into a \ac{TCP} stream interface using SiTCP.

\subsubsection{Control, Configuration and Monitoring}

The firmware includes many registers containing information allowing full telemetry of the hardware from software. The \ac{CCM} aspect of the firmware's job is to monitor and set these registers. The design of the \ac{PAT} cards allows the \ac{FPGA} to also monitor and change the behaviour of several other components on the board using standardised slow control interfaces (\textit{e.g.} \ac{I2C}, \ac{SPI}, \ac{PMbus}). The list of components can be found in \cref{tab:SlowControlList} along with a short explanation of their function.

\begin{table}[h]
    \centering
    \caption{List of \ac{PAT} card components that can be addressed through slow control.}
    \label{tab:SlowControlList}
    \begin{tabular}{c|c|c}
        \textbf{Component} & \textbf{Interface} & \textbf{Application} \\
        \hline
        \hline
        \ac{FEC} \ac{DC/DC} Converter & \ac{PMbus} & Power converter for \ac{FEC}\\
        SAMPA slow control & \ac{I2C} & Configuration of SAMPA device \\
        \ac{SFP} Connectors & \ac{I2C} & Ethernet I/O\\
        Clock Synthesizer & \ac{I2C} & Clock cleaning \\
        Flash Memory & \ac{SPI} & Remote firmware programming\\
        EEPROM & \ac{I2C} & Design information\\
        \ac{I2C} Switches & \ac{I2C} & Slow control routing\\
        Port-Expanders & \ac{I2C} & Power switches for \ac{FEC}\\
        \hline
    \end{tabular}
\end{table}

Each detector operation mode requires a specific configuration of these components. For example, with the inclusion of the \ac{I2C} port-expanders, individual \ac{FEC}s can be turned off and removed from the run if there is a problem with their operation. 

In addition to the external hardware components, the \ac{CCM} part of the firmware implements a variety of standard IPbus registers and memories dedicated to monitoring the data flow or configuring operational modes. A list of these registers can be seen in \cref{tab:IPbusList}.

\begin{table}[h]
    \centering
    \caption{IPbus configuration and monitoring register list.}
    \label{tab:IPbusList}
    \begin{tabular}{c|m{12em}|c}
        \textbf{Reg. Name} & \textbf{Control/Monitor} & \textbf{Number of Registers} \\
        \hline
        \hline
        Clock control & Enable clock transmission to fanout chips & 1/\ac{FEC} group \\
        Clock rate control & Control over internal clock multiplexers for (\SI{320}{MHz}, \SI{160}{MHz} or \SI{80}{MHz}) & 1/\ac{PAT} \\
        Serial links control & Ethernet serial link monitor and configuration & 1/Serial Link \\ 
        Sync. control & Send sync. signal to all \ac{FEC} sampling windows & 1/\ac{FEC} group \\
        Memory content & Live data flow memory content & 1/\ac{FEC} \\
        \ac{RR} control & Enable, Veto and Reset of \ac{RR} logic & 1/\ac{RR} block \\
        Logic Reset & Reset signals for all logic components & 1/Firmware block in the design \\
        \hline
    \end{tabular}
\end{table}

\subsection{Software development} \label{SoftwareDevelopment}

Operating the \ac{PAT}-based readout requires software as well as the custom firmware.  The \ac{PAT} cards have three main functions: data flow, \ac{CCM}, and timing. Each function has a software counterpart that has been developed for the \ac{TOAD} test beam and designed so as to minimise modifications needed for integration in the final \ac{ND-GAr}. All three sets of software are designed to run on standard commercial servers.

The data flow application is based on the \ac{DUNE}~\ac{DAQ} software suite. For \ac{CCM}, an IPbus-based framework has been developed, allowing full access to all the in-pressure electronics' configuration registers. Finally, for timing, the software is based on the framework from the \ac{DUNE} timing system, which we have wrapped in the bespoke \ac{PAT} \ac{CCM} framework since the timing system uses IPbus for configuration as well. 

\subsubsection{DUNE DAQ integration}

Once the data has been streamed out, it is received by the \ac{DUNE}~\ac{DAQ} software. While this software already existed prior to this work, significant effort was required to integrate \ac{TOAD}'s \ac{DAQ} into the larger \ac{DUNE} software stack. A schematic of the general data flow within \ac{DUNE}~\ac{DAQ} can be found in \cite{Abi_2020}. The integration of the \ac{TOAD} readout into this framework involved adding the Ethernet packet structure described in \cref{sec:dataflow} into the upstream \ac{DAQ}. Additionally, the zero-suppression functionality of the \ac{FEC}s allows for the \ac{PAT} boards to only output data above the threshold we wish to be considered for offline processing. This means that \ac{TOAD} also demonstrated that the \ac{DUNE}~\ac{DAQ} software framework worked in untriggered continuous readout mode. 

A further difference between \ac{TOAD} and the other systems using \ac{DUNE}~\ac{DAQ} is that the output stream data format allows each hit that is streamed out to contain a variable number of \ac{ADC} samples, while all other systems had fixed numbers of samples per transmitted data packet. This implementation expands the capabilities of the \ac{DUNE}~\ac{DAQ} software and allows it to handle variable-length data packets \cite{a.abedabudTriggerDataAcQuisition2023}. 

\subsubsection{Control, Configuration and Monitoring software suite}
\begin{figure}[h]
    \centering
    \includegraphics[width=0.9\linewidth]{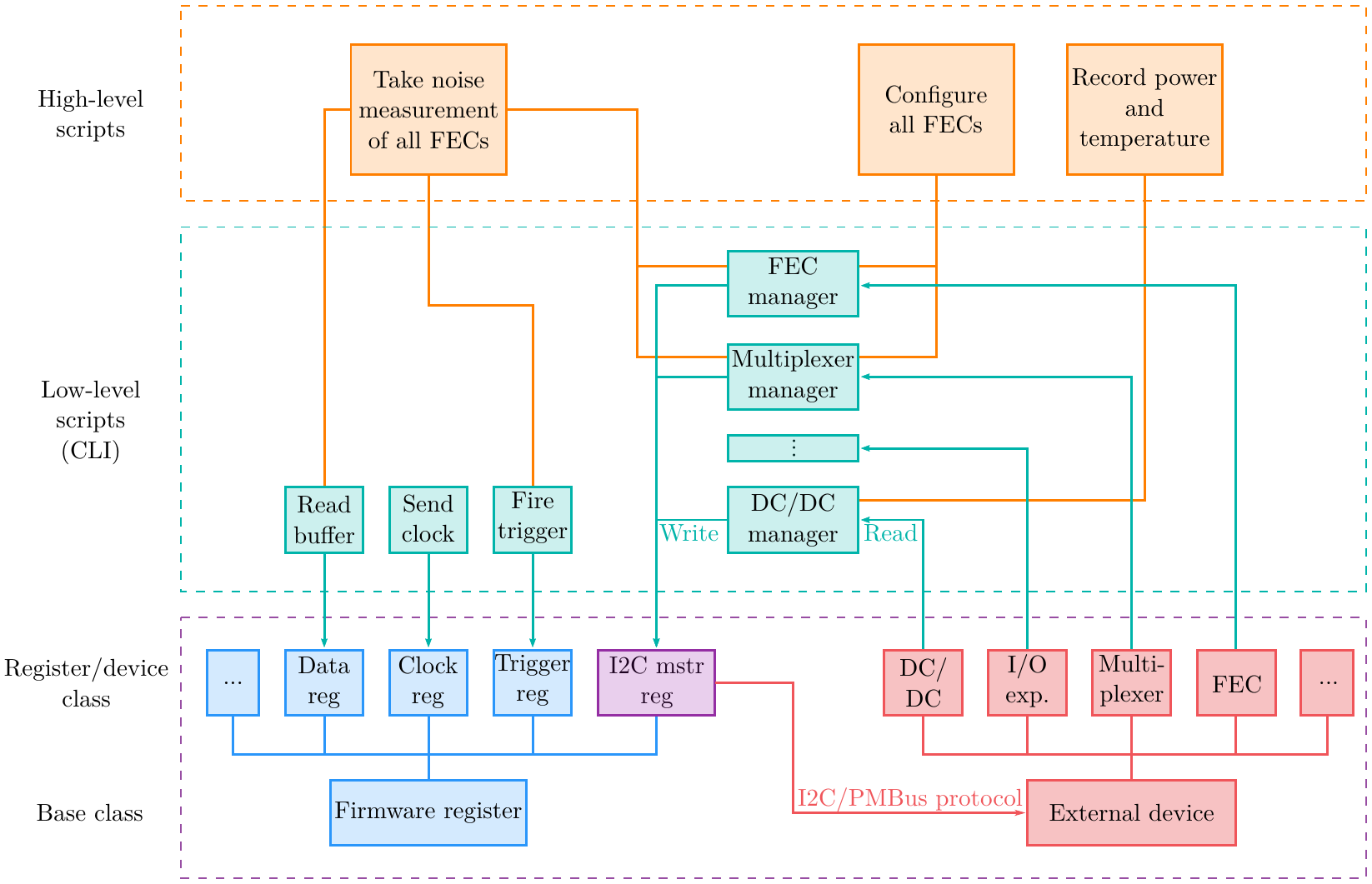}
    \caption{Block diagram for the \ac{CCM} suite. Orange blocks at the top represent high-level scripts that interact with various low-level scripts to fulfil certain tasks. Low-level scripts are shown as green blocks, they provide \acs{CLI} for users to interact directly with individual components on the \ac{PAT} boards. Individual firmware registers and external devices are implemented as separate classes that inherit from the corresponding base class. Firmware registers are shown in blue and external devices are shown in red. The \ac{I2C} master register in purple facilitates the \ac{I2C} communication to external devices. }
    \label{fig:slow-control}
\end{figure}

The data from \ac{TOAD} is received by standard \ac{DUNE} software \cite{a.abedabudTriggerDataAcQuisition2023}. However, the \ac{CCM} software suite is a new, scalable and modular software framework. The framework has two key base classes to talk to the two classes of objects on the \ac{PAT} boards: firmware-implemented registers and external devices. 
The framework is summarised in \cref{fig:slow-control}. 
Each base class contains general methods needed by all registers of that type. 
Communication to the firmware registers is via IPBus, so the base class contains IPBus-specific methods, while the inherited classes for each specific register contain their own specific bit maps that relate their 32-bits to specific functions. 
This scheme allows scalability to all electronic devices that use IPBus and similar firmware implementation.

The external device base class contains methods that generate instructions for the \ac{FPGA} to access registers on external devices. 
For a device that uses the \ac{I2C} interface, the inherited class will contain an additional \ac{I2C} address and a list of register addresses for the different registers inside the device. 
Additional slow control protocols, such as \ac{SPI} or \ac{PMbus}, are integrated in a similar manner.

The \ac{CCM} suite also provides various low-level scripts that act as a \ac{CLI} for the user to manage the hardware. 
By combining multiple low-level scripts, one can create high-level scripts that are tailored towards specific use cases. 
Examples include the noise measurements outlined in \cref{sec:noise} and continuous temperature and power monitoring used in \cref{sec:thermal}.

\section{Results}
\subsection{Noise characterisation}\label{sec:noise}

\begin{figure}
  \begin{center}
    \includegraphics[width=0.7\textwidth]{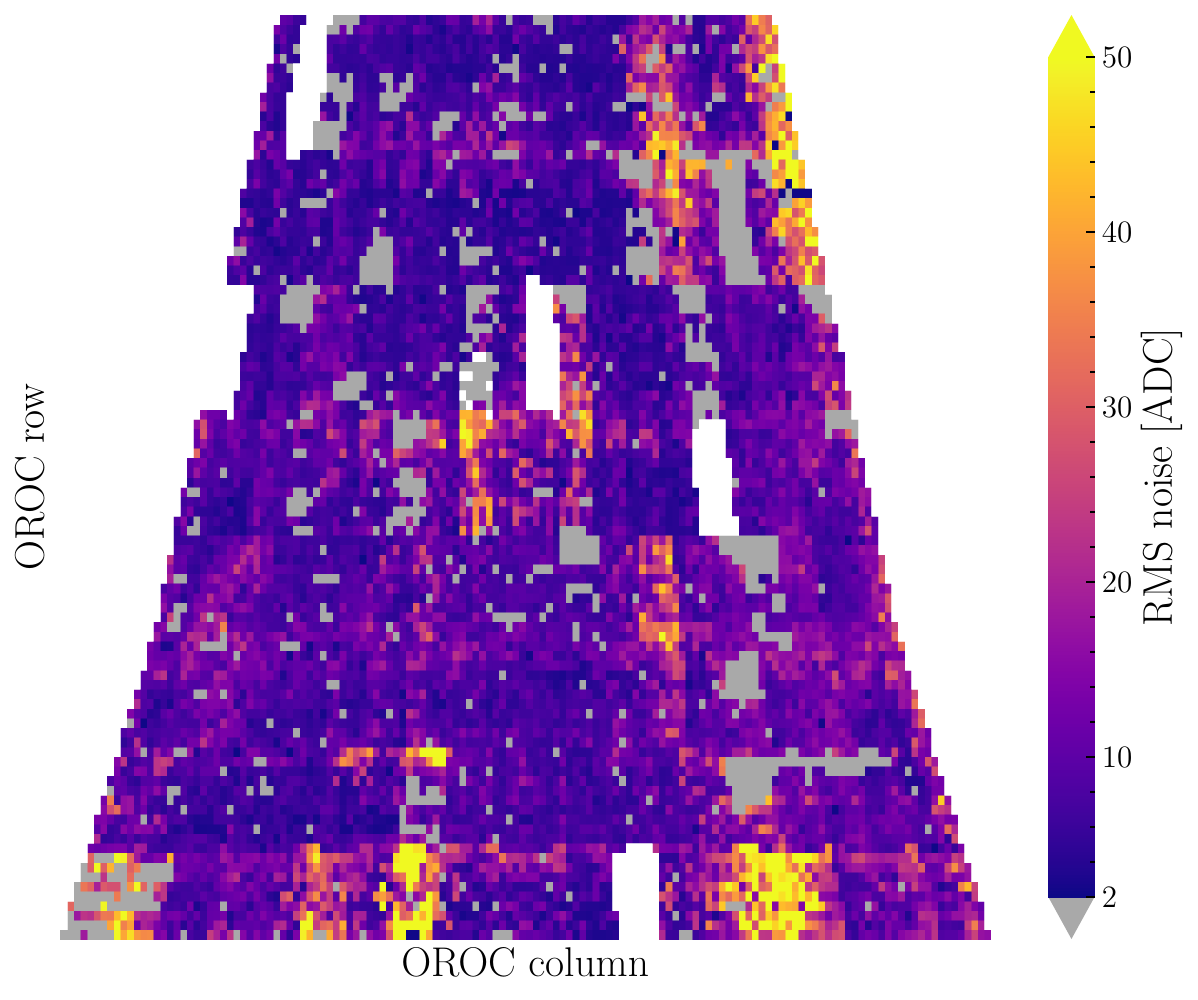}
  \end{center}
  \caption{A noise map of the \ac{OROC} pads in \SI{4.5}{barA} pressure. The noise calculated from each channel on each \ac{FEC} is mapped to the spatial position within the \ac{OROC} and the \acs{RMS} noise in \acs{ADC} counts is plotted. The white regions are caused by non-responding \ac{FEC}s.}
  \label{fig:noise_map}
\end{figure}

\begin{figure}
  \begin{center}
    \includegraphics[width=0.7\textwidth]{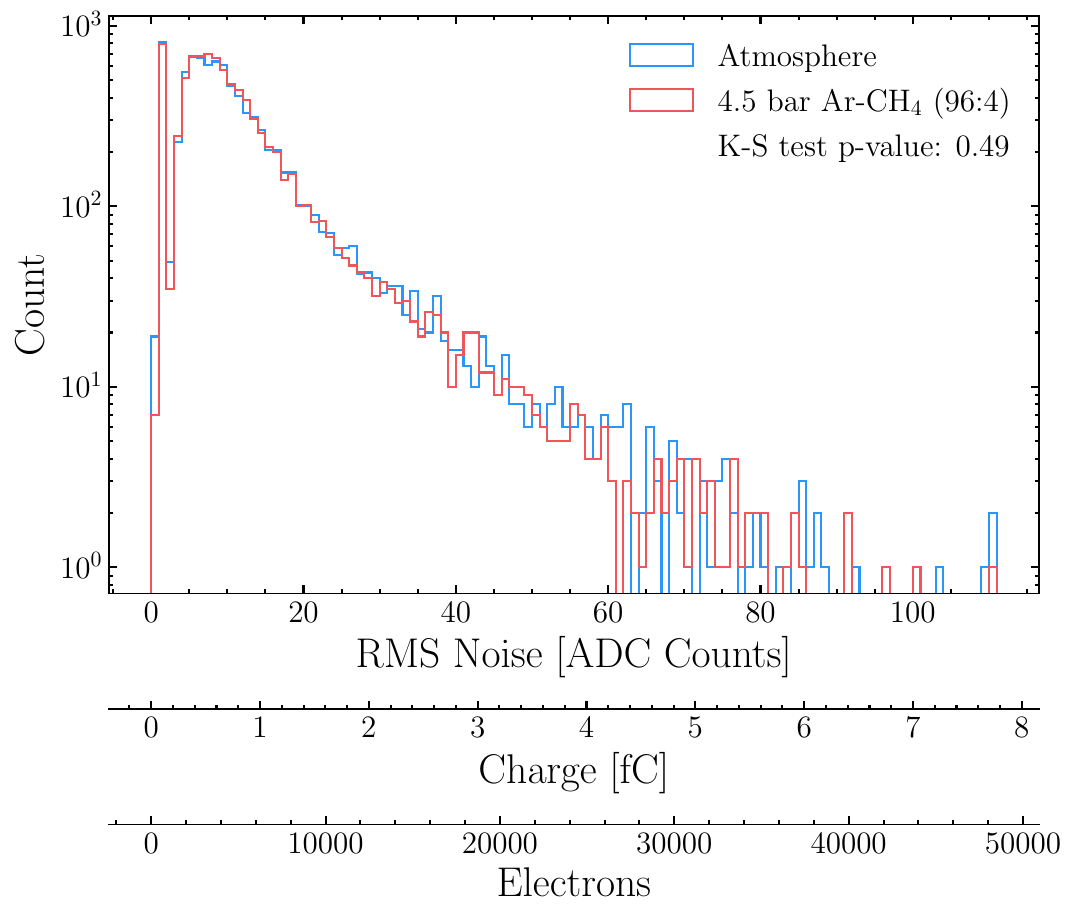}
  \end{center}
  \caption{Distributions of \acs{RMS} noise for all valid channels across all \ac{FEC}s in mode 1 (blue) and 2 (red) of \cref{tab:gasmixtures}.}
  \label{fig:noise_hist}
\end{figure}

Understanding the noise profile is crucial in verifying that we have a sufficient signal-to-noise ratio to identify the signals we are looking for. 
We characterise noise as the \acf{RMS} error of the waveform in the absence of any signal. Prior to measuring the noise and in order to ensure that everything is correctly set up, the \ac{PAT} cards have been tested for synchronisation. This has been achieved on the table top setup, where the clock skew between the different \ac{FEC} links has been measured and found to be in the sub-ps regime, indicating that all the \ac{FEC}s were clocked at the same time. The same test with the same outcome has been performed for the trigger signal, which is crucial for collecting noise data.

Noise data were collected by sending trigger signals to the \acp{FEC} using the high-level script shown in \cref{fig:slow-control}. The noise is measured by taking the \ac{RMS} of a \SI{25}{\micro s} waveform, which corresponds to 500 samples at \SI{20}{MHz}, for each channel in each \ac{FEC}. 
\Cref{fig:noise_map} shows the spatial distribution of the \ac{RMS} noise of all \ac{FEC}s in the \ac{OROC}, measured under \SI{4.5}{barA} pressure. Channels with \ac{RMS} noise less than \SI{2}{ADC} are coloured grey, while channels with noise above \SI{50}{ADC} are shown with the highest colour in the scale. The blank regions are caused by \ac{FEC}s that did not respond to any \ac{I2C} communications.

\begin{figure}
  \begin{center}
    \includegraphics[width=0.7\textwidth]{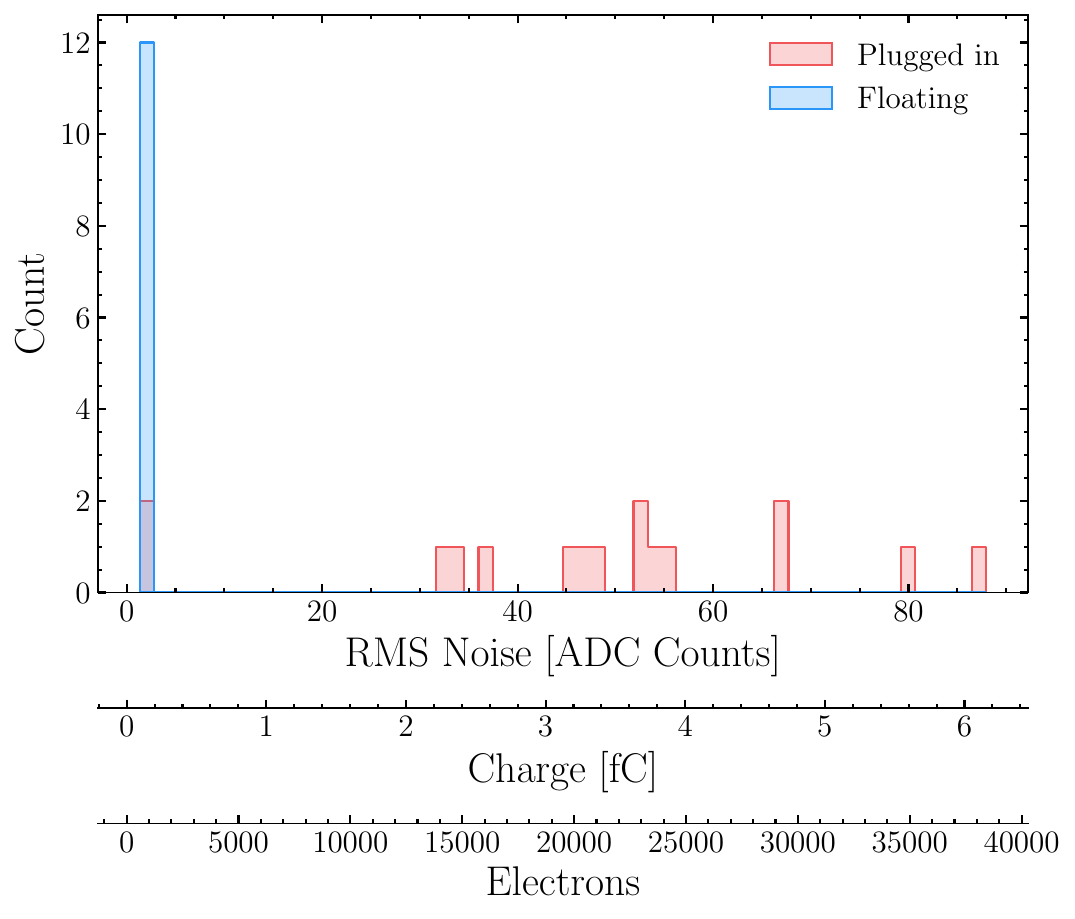}
  \end{center}
  \caption{Distribution of \ac{RMS} noise on a subset of 12 channels of a noisy \ac{FEC} before (red) and after (blue) it is removed from the \ac{OROC}.}
  \label{fig:fec_in_out}
\end{figure}

\Cref{fig:noise_hist} plots the distribution of the \ac{RMS} noise. Noise from a total of 9600 (9656) channels was measured for the atmospheric pressure (\SI{4.5}{barA}) case, which corresponds to $\sim$96\% of the total number of \ac{OROC} pads as described in \cref{tab:orocdimensions}. 
We require the waveform baseline to be greater than \SI{10}{ADC} counts in \cref{fig:noise_hist}. 
This cut is placed to remove waveforms that have noise cut off at \SI{0}{ADC} due to the low baseline, which makes the waveform appear to have low noise. 
This cut removes $10\%$ of the data, and results in 8621 (8645) channels for the atmosphere (\SI{4.5}{barA}) case. 
Correction of the baseline can be applied to individual SAMPA channels, which would mitigate the issue of low baseline and therefore recover these channels. 
This calibration was not performed during this operation period of \ac{TOAD} due to time constraints. 

The noise distribution is bimodal with a single-bin peak at just above 0 and another peak at $\sim$\SI{10}{ADC}. 
The first peak is caused by \ac{FEC}s that are not properly connected to the \ac{OROC}. 
This peak is spatially shown as the grey patches in \cref{fig:noise_map}.
This is demonstrated by measuring the \ac{RMS} noise from the \ac{FEC}s after they are removed from the \ac{OROC} and left floating. 
\Cref{fig:fec_in_out} shows a comparison between the \ac{RMS} noise for the same subset of 12 channels on one \ac{FEC}, before and after the card is unplugged from the \ac{OROC}. 
When the \ac{FEC} is floating, all of the measured channels have \ac{RMS} noise of $<\SI{5}{ADC}$. 
This level of noise corresponds to the intrinsic noise level of the SAMPA ($\sim$\SI{1600}{electrons}), as shown in \cref{tab:SAMPAParameters}, which is equivalent to \SI{0.26}{fC} and \SI{3.6}{ADC}. A gain of \SI{14}{ADC/fC} is used for the conversion \cite{adolfssonSAMPAChipNew2017}.
For most of the \ac{FEC}s that are well-connected to the OROC, their noise distribution is characterised by the second peak. 

The large tail that extends past \SI{100}{ADC} in \cref{fig:noise_hist} is shown as distinct noisy regions in \cref{fig:noise_map}. This is due to a partial connection between the \ac{FEC} and the \ac{OROC}, which results in a floating ground, as this was not observed in benchtop tests with direct connection to the \acp{FEC}. 
The difficulty in connecting the \ac{FEC} to the \ac{OROC}, for both the case of improper connection, which left the \ac{FEC} afloat, and partial connection, which resulted in a floating ground, is due to the legacy \ac{OROC} connectors used in \ac{TOAD}. 
It is observed that the same \ac{FEC} can gain good performance when moved to a different slot on the backplane. 
It is also worth noting that the \ac{ALICE} \acp{ROC} are now end of life, so a future \ac{HPgTPC} would be designed so as to achieve a more reliable connection to the front-end electronics.

From \cref{fig:noise_hist}, no noticeable difference can be seen in the noise measurements taken in atmosphere (mode 1 of \cref{tab:gasmixtures}) and \SI{4.5}{barA} 96:4 Ar-CH$_4$ (mode 2). A \ac{K-S} p-value of $0.49$ is obtained from the unbinned data, which suggests the same underlying distribution.

\begin{figure}
  \begin{center}
    \includegraphics[width=0.7\textwidth]{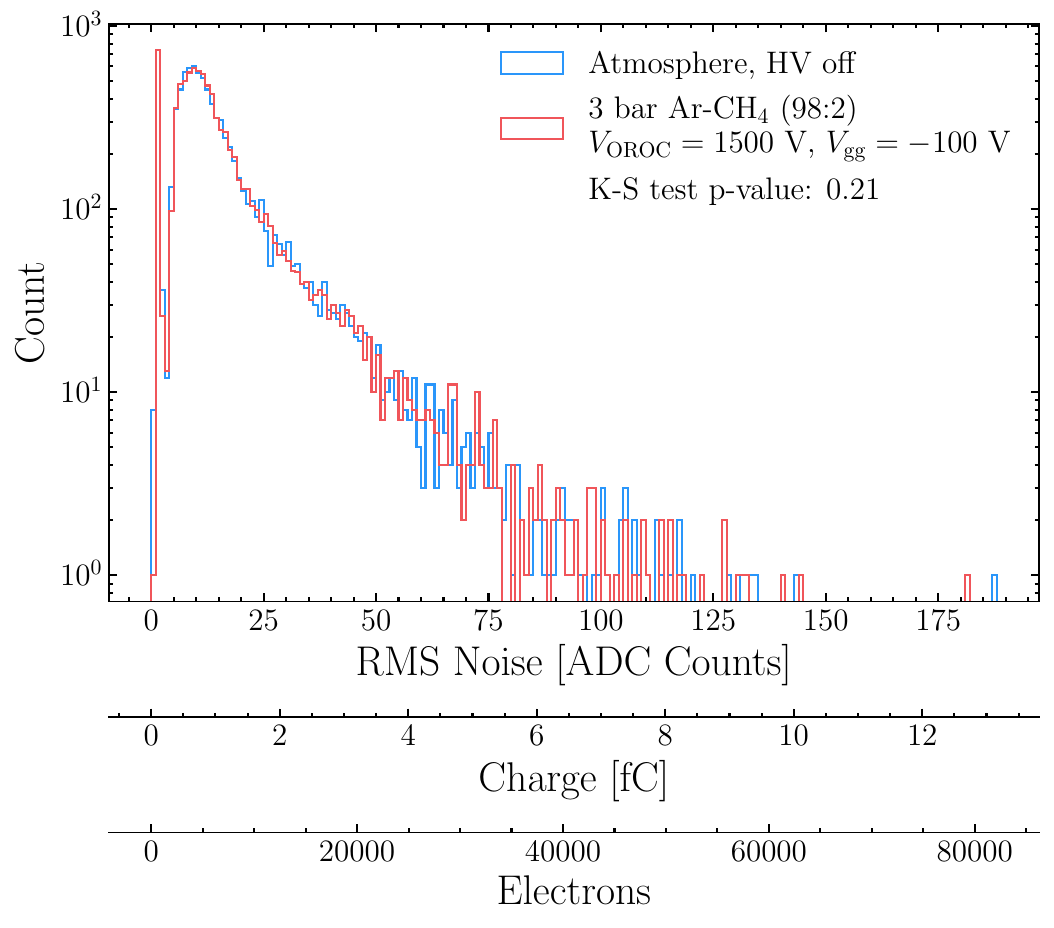}
  \end{center}
  \caption{Distributions of \ac{RMS} noise for all valid channels across all \ac{FEC}s in mode 1 (blue) and 3 (red) of \cref{tab:gasmixtures}. Data for this plot were taken in a different condition and before some noise improvements were made compared to \cref{fig:noise_hist}, so no direct comparison could be made.}
  \label{fig:hv_hist}
\end{figure}

\begin{figure}
    \centering
    \includegraphics[width=0.7\linewidth]{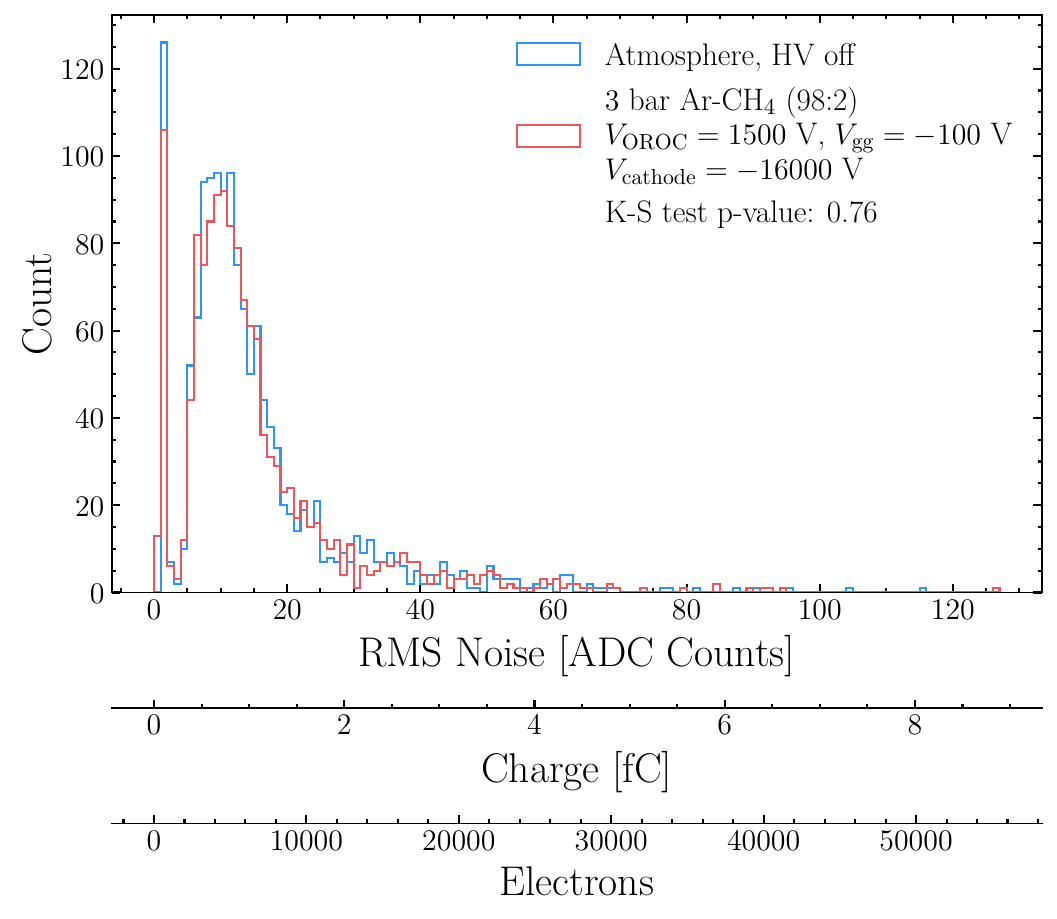}
    \caption{Distributions of \ac{RMS} noise for a subset of channels across all \ac{FEC}s in mode 1 (blue) and 4 (red) of \cref{tab:gasmixtures}. A linear scale is used due to the smaller sample size.}
    \label{fig:hv_cathode_hist}
\end{figure}

Additionally, we see that powering of the \ac{HV} devices does not have a noticeable effect on the noise. 
This can be seen in \cref{fig:hv_hist} which compares the noise when the \ac{HV} devices are off in atmosphere (mode 1 of \cref{tab:gasmixtures}) and when $V_\text{OROC}=\SI{1500}{V},V_\text{gg}=\SI{-100}{V}$ in 98:2 Ar-CH$_4$ (mode 3).
A \ac{K-S} p-value of $0.21$ is obtained and again suggests the same underlying distribution.
The data for this plot were taken in a different condition and before some noise improvements were made to address the known issue of poor grounding in the test beam facility. As such, no direct comparison could be made with \cref{fig:noise_hist}. 
Nevertheless, the bimodal distribution and the positions of the peaks are still the same between the two sets of data. 
We also tested mode 4 (same \ac{OROC} and gating grid voltages and  \SI{-16}{kV} cathode voltage) for a small subset of the channels, shown in \cref{fig:hv_cathode_hist}. 
The same data for mode 1 (blue) was used for both \cref{fig:hv_hist} and \cref{fig:hv_cathode_hist}, but for the latter figure, only the subset of channels matching those collected for mode 4 were shown. A K-S p-value of 0.76 is obtained and the main features in the noise distribution remained identical. 
 
\begin{figure}
  \begin{center}
    \includegraphics[width=0.7\textwidth]{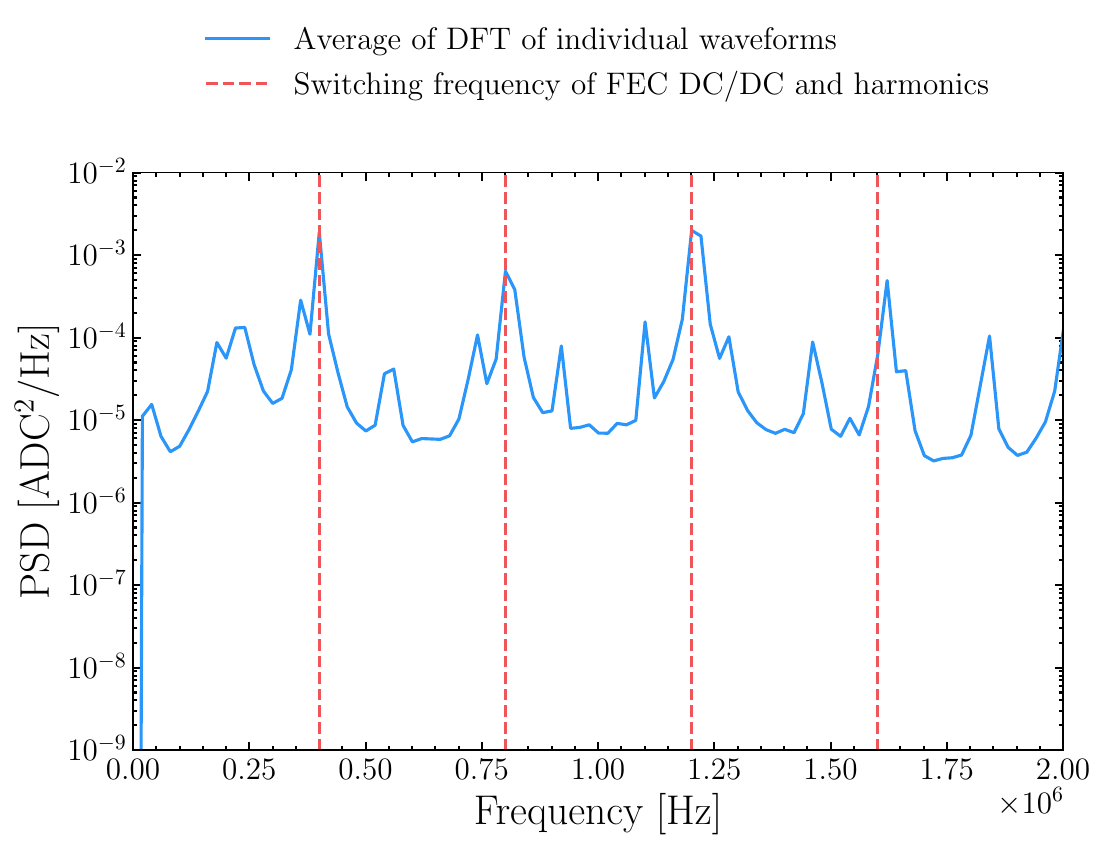}
  \end{center}
  \caption{Power spectral density of a single channel on a \ac{FEC}. The average of all $1000$ \acs{DFT} of the waveforms is shown in blue. Dashed lines show the switching frequency of the \ac{FEC} \ac{DC/DC} and its harmonics.}
  \label{fig:fft}
\end{figure}

The contribution of noise from different frequencies can be studied through \acf{DFT} of the waveforms. 
\Cref{fig:fft} shows the \ac{DFT}s from a single channel on a \ac{FEC}. 
To avoid random noise, 1000 waveforms are collected and the average of the \ac{DFT} of 1000 individual waveforms is plotted. 
Large peaks at \SI{400}{kHz} and integer multiples of it can be noticed in \cref{fig:fft}. 
This corresponds to the switching frequency of the \ac{FEC} \ac{DC/DC} converter.
Less prominently, the smaller peaks at multiples of \SI{200}{kHz} correspond to the switching frequency of the main \ac{DC/DC} converter.

To summarise, the \ac{DC/DC} components on the \ac{PAT} boards are major contributors to the noise measured by the \ac{FEC}s. However, since this iteration of the board is not the one expected to be deployed in ND-GAr in a future revision, more care will be taken with regard to the choice of the \ac{DC/DC} converter to not limit the physics capability of the detector. 
In addition, \ac{FEC}s that were not conductively connected to the \ac{OROC} would appear to have extremely low noise. 
High noise channels in the \ac{OROC} are seen in small regions, usually corresponding to a single \ac{FEC}, and could be due to improper grounding of the \ac{OROC} connectors.

To assess whether the level of noise measured in \ac{TOAD} would be sufficient for a neutrino experiment, we estimate the size of the signal on a \ac{FEC} that could be generated by a minimally ionising particle. 
The expected signal-to-noise ratio can be calculated using stopping power \SI{1.5}{MeV.cm^2 / g}, atmospheric argon gas density \SI{1.66e-3}{g/cm^3}, and ion pair production energy \SI{26}{eV}, giving \SI{95}{e^-/cm} at atmosphere \cite{pdg}. 
At \SI{10}{bar}, using a projected gas gain of \num{10000} from \cite{Ritchie-Yates2023-cs} and accounting for the change in gas density, one obtains $\SI{9500000}{e^-/cm}=\SI{1520}{fC/cm}$. We take \SI{1}{cm} as an approximate track length across an \ac{OROC} pad, \textit{i.e.}, a channel in a \ac{FEC}. 
Simulations have been done to show that the electrons spread across multiple pads, with only 22\% lie within \SI{0.5}{cm} of the centre. Furthermore, it takes $\sim$7 samples for the charge to be fully deposited. So using \SI{14}{ADC/fC} from \cite{adolfssonSAMPAChipNew2017}, on average, a minimally ionising particle is expected to deposit \SI{668}{ADC} per channel per sample. 
The 95 percentile of \cref{fig:noise_hist} gives a conservative estimate for noise of \SI{32}{ADC}, which leads to an estimated signal-to-noise ratio of $\sim$21. 

\subsection{Thermal profile}\label{sec:thermal}
\begin{figure}[h]
        \centering
        \includegraphics[width=0.7\linewidth]{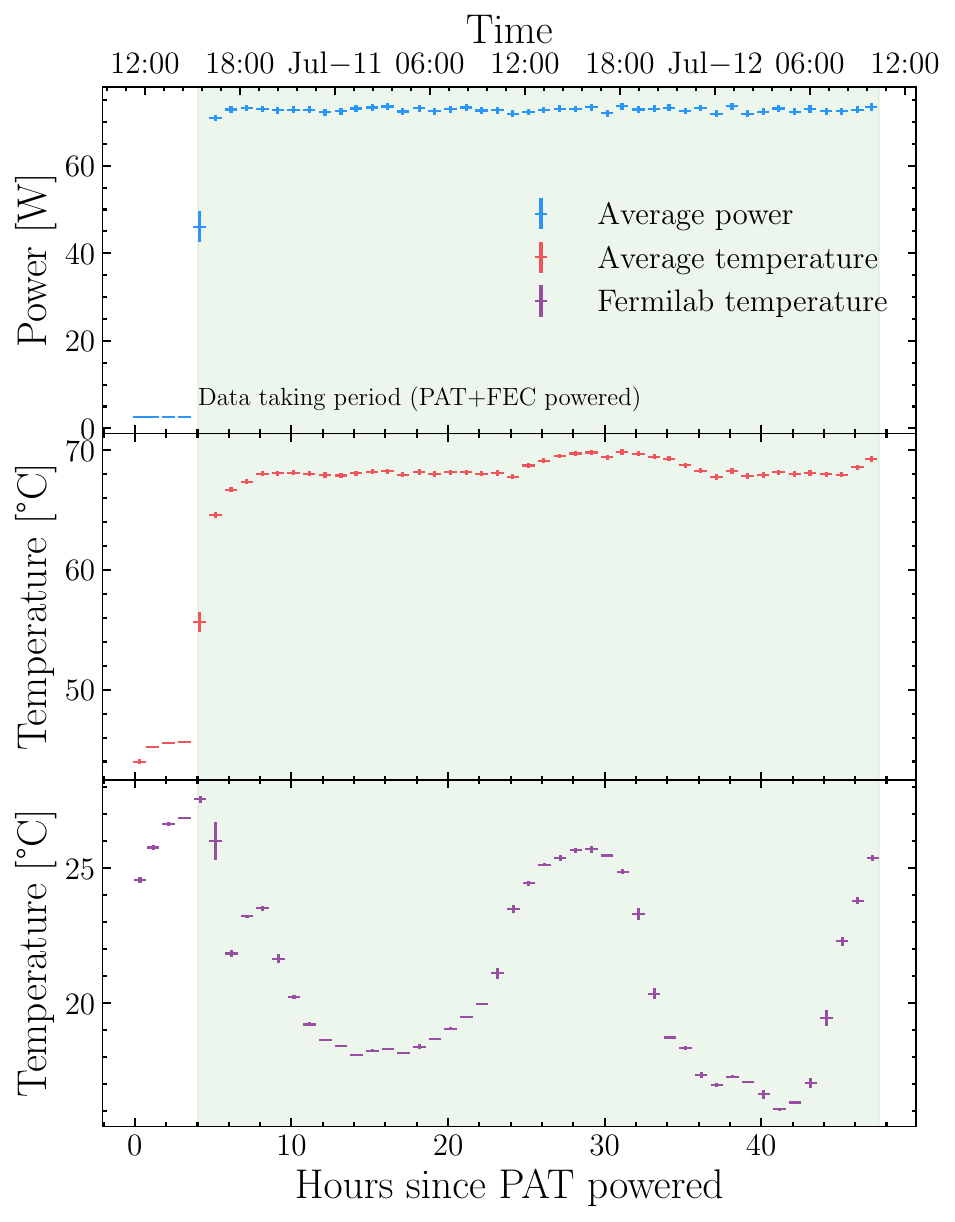}
        \caption{Power output (blue) and temperature (red) of the \ac{FEC} \ac{DC/DC} converter since the \ac{PAT} boards were powered in \SI{4.5}{barA} pressure, averaged across the three \ac{PAT} boards. The error bar is mainly due to the difference across the three boards. The temperature measured from a Fermilab weather station is plotted in purple. The slight increases in \ac{PAT} temperature at noon on both 11 Jul and 12 Jul correlate with high environmental temperature. The green region is the data taking period where both the \ac{PAT} boards and \ac{FEC}s are in operation.}
        \label{fig:temperature}
\end{figure}

\begin{figure}[h]
    \centering
    \includegraphics[width=0.7\linewidth]{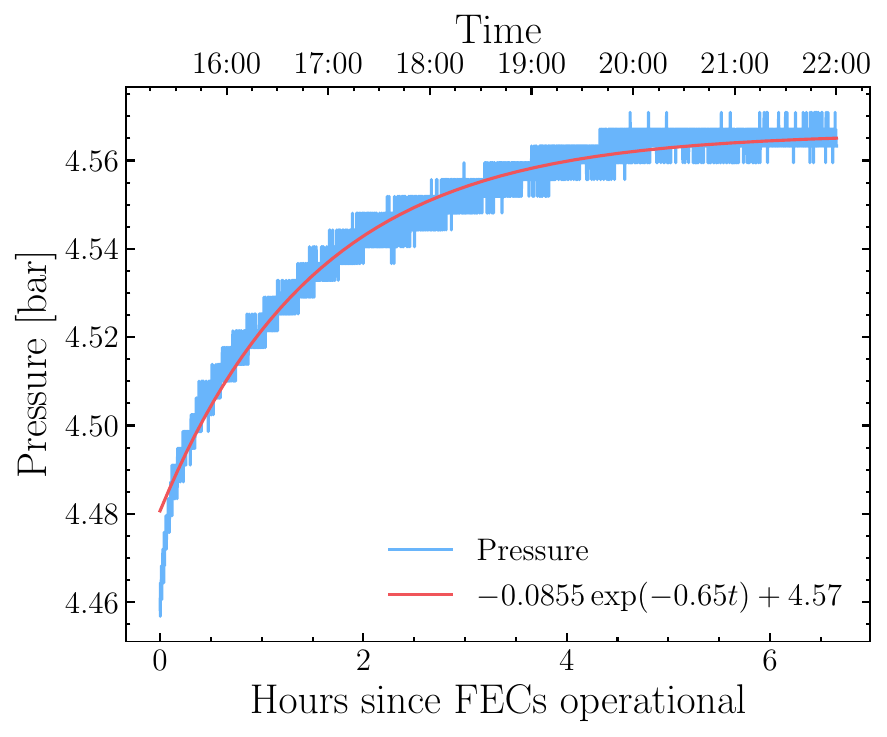}
    \caption{Pressure in \si{barA} since the \ac{FEC}s have been powered (blue), fitted with an exponential function (red). The pressure measurement is limited by the manometer resolution of \SI{0.0038}{barG}.}
    \label{fig:pressure}
\end{figure}

To achieve a reasonable gain in a gas \ac{TPC}, a stable pressure must be maintained. The thermal profile of the system thus needs to be well-understood as in a closed vessel the temperature directly affects the pressure. 
Cooling is currently achieved in \ac{TOAD} through thermal contact between the aluminium mounting plate of the \ac{PAT} board and the \ac{OROC}. 
The \ac{FEC}s are cooled through contact with the gas in the vessel. 
The temperature and the power output from each \ac{PAT} board are recorded via the \ac{FEC} \ac{DC/DC} converter on each board, and the average values in \SI{4.5}{barA} are plotted in \cref{fig:temperature}. 
The idle power consumption, with no power delivered to the \ac{FEC}s, for the \ac{PAT} boards all connected to the same power bus, ranges between 0.89-\SI{3.60}{W}. Additionally, the idle temperature in equilibrium with the environment has been found to be \SI{40}{\celsius}, which is well within the operational margin provided by AMD \cite{AMDDatasheetKU}.

The \ac{FEC}s constitute the majority of the power consumption. With all available \ac{FEC}s in full operation, the power consumption for each \ac{PAT} board ranged between $[60.7,64.8]\si{W}$, $[74.0,75.4]\si{W}$, and $[79.7,81.6]\si{W}$ for the \ac{PAT} board with $43$, $53$, and $55$ \ac{FEC}s connected, respectively.
Assuming the power consumption scales linearly with the number of \ac{FEC}s, an average power consumption of \SI{1.4}{W} per \ac{FEC} can be calculated. This corresponds to a maximum power consumption of $\sim$\SI{85}{W} per \ac{PAT} board. 

Relying solely on the thermal coupling between the \ac{PAT} boards and the \ac{OROC}, an average change of temperature of \SI{23}{\celsius} on the \ac{DC/DC} converter was observed as a result of the operation of the \ac{FEC}s. 
While the final temperature reached is within the operation threshold of the electronics, a large temperature change and use of the gas as the cooling medium can lead to a long stabilisation time for the pressure in the vessel as well as uncertain gas drift properties. 
As shown in \cref{fig:pressure}, by fitting an exponential to the pressure profile, it takes $\sim$\SI{6}{hours} for the pressure to have a rate of change of pressure $\dot{P}<\SI{1}{mbar\per hr}$. 
It is also worth noting that the heavy pressure vessel is in contact with the gas, increasing the time to heat up. Furthermore, the \ac{OROC} is also in thermal contact with the \ac{TOAD} vessel and the gas, further increasing the pressure stabilisation time and also exposing it to the high equilibrium temperature. 
Such a long stabilisation time prolongs the time taken to go from the detector being off to stable operations, increasing downtime.

Each \ac{PAT} board is contributing a maximum of only $\sim$\SI{85}{W} of power which is well within the range that can be mitigated by the addition of active cooling or a better heat sink design. This investigation of the thermal profile is a necessary component of future research. 

Stable performance of the electronics was observed during the pressure operation. During the time-constrained test-beam period, we observed stability in pressure for the duration of multiple days, with no significant variation in the electronics performance observed. Similar stability was observed in atmosphere during the test-beam period, as well as throughout benchtop testing.

\section{Conclusion}
Many readout systems developed for particle physics experiments target the high-occupancy environment of a hadron collider experiment. Scaling systems such as readout electronics from sPHENIX or \ac{ALICE} would be unnecessarily costly. Similarly, technology such as LArPix would require overclocking to meet the resolution requirements for \ac{ND-GAr}. Commercial solutions, such as CAEN, would also be considerably costly. We have, instead, described a custom system designed for much lower occupancy detectors, such as those in neutrino physics experiments. The system leverages the lower buffering requirements to use far fewer \ac{FPGA}s per channel and thereby significantly reduces the cost of the system. Even though it is a prototype system manufactured at a small scale, the readout for \ac{TOAD} (including \ac{FEC}s, \ac{PAT} cards and all the hardware to connect and power them) costs less than \$\num{20000} for $\sim$\num{10000} channels. Scaling this up to the full \ac{DUNE} Phase-II \ac{ND}, ignoring likely economies of scale, gives a conservative cost estimate of $\sim$\$1.4 million.

The \ac{ND-GAr} detector that this system was designed for requires operation in a magnetic field and at high pressure (up to \SI{10}{barA}). The high-pressure requirement also necessitates a limited number of cables exiting the pressure vessel. We have described the designed system and demonstrated its operation at a range of pressures and gas mixtures up to \SI{4.5}{barA}. These tests were carried out through the \ac{TOAD} experiment and involved the readout of $\sim$\num{10000} channels of an \ac{MWPC} from the \ac{ALICE} \ac{TPC}.

The system performed well, with stable operation seen throughout the test period, which included multiple days of running at atmosphere and pressure as well as benchtop testing.
However, the passive cooling used in the test has been identified as an area that must be improved before final deployment. Nevertheless, the total heat load is well within the range that can be dealt with using active cooling. The noise characteristics of the system were measured and found to be within the range that will allow us to read out the intended signals for almost all channels. We have identified poor connection due to the mounting of the \ac{FEC}s in the legacy connectors of the \ac{OROC} as the primary item contributing to channels that are not readable. This is specific to the \ac{OROC} used in \ac{TOAD}, and is not expected to be a problem for future \ac{HPgTPC}, \textit{e.g.} \ac{ND-GAr}, which will use different connectors. Notwithstanding these necessary minor modifications, we have shown that this system is able to fulfill the needs of a high-pressure gas \ac{TPC} detector for a fraction of the price of systems used for \ac{LHC} type detectors.

\acknowledgments
We would like to thank the STFC, Royal Society and UKRI, UK; the DOE, USA; and the Imperial College President's Scholarship scheme for funding this work. Particularly, for Fermilab, this document was prepared using the resources of the Fermi National Accelerator Laboratory (Fermilab), a U.S. Department of Energy, Office of Science, Office of High Energy Physics HEP User Facility. Fermilab is managed by Fermi Forward Discovery Group, LLC, acting under Contract No. 89243024CSC000002.

We would also like to acknowledge the invaluable assistance of the mechanical and electrical workshops at Royal Holloway University of London and Imperial College London, as well as the staff at the Fermilab test beam facility.


\bibliographystyle{JHEP}
\bibliography{biblio.bib}

\end{document}